\documentclass[journal]{vgtc}                

\ifpdf
  \pdfoutput=1\relax                   
  \usepackage{graphicx}                
  \DeclareGraphicsExtensions{.pdf,.png,.jpg,.jpeg} 
\else
  \usepackage{graphicx}                
  \DeclareGraphicsExtensions{.eps}     
\fi%

\graphicspath{{figures/}{pictures/}{images/}{./}{fig/}} 

\usepackage{microtype}                 
\PassOptionsToPackage{warn}{textcomp}  
\usepackage{textcomp}                  
\usepackage{mathptmx}                  
\usepackage{times}                     
\usepackage{cite}                      
\usepackage{tabu}                      
\usepackage{booktabs}                  

\usepackage{color}
\usepackage{tikz}
\usepackage{transparent}
\usepackage{comment}
\usepackage{amsfonts}
\usepackage{enumitem}
\usepackage[normalem]{ulem}
\usepackage{colortbl}
\usepackage{subfigure}
\usepackage{graphicx}
\usepackage{multirow}
\usepackage{lettrine}

\usepackage{amssymb}

\onlineid{1054}

\vgtccategory{Research}
\vgtcpapertype{Theory/Model}

\title{Modeling the Influence of Visual Density on Cluster Perception in Scatterplots Using Topology}

\author{Ghulam Jilani Quadri and Paul Rosen}
\authorfooter{
\item
 Ghulam Jilani Quadri is with the University of South Florida. E-mail: ghulamjilani@usf.edu.
\item
 Paul Rosen is with the University of South Florida. E-mail: prosen@usf.edu.
}

\shortauthortitle{Quadri and Rosen: Multi-Factor Modeling of Cluster Perception}

\abstract{
      Scatterplots are used for a variety of visual analytics tasks, including cluster identification, and the visual encodings used on a scatterplot play a deciding role on the level of visual separation of clusters. For visualization designers, optimizing the visual encodings is crucial to maximizing the clarity of data. This requires accurately modeling human perception of cluster separation, which remains challenging. We present a multi-stage user study focusing on four factors---\textit{distribution size of clusters}, \textit{number of points}, \textit{size of points}, and \textit{opacity of points}---that influence cluster identification in scatterplots. From these parameters, we have constructed two models, a distance-based model, and a density-based model, using the merge tree data structure from Topological Data Analysis. Our analysis demonstrates that these factors play an important role in the number of clusters perceived, and it verifies that the distance-based and density-based models can reasonably estimate the number of clusters a user observes. Finally, we demonstrate how these models can be used to optimize visual encodings on real-world data.
}


\teaser{
    \centering
  
    \includegraphics[height=179pt]{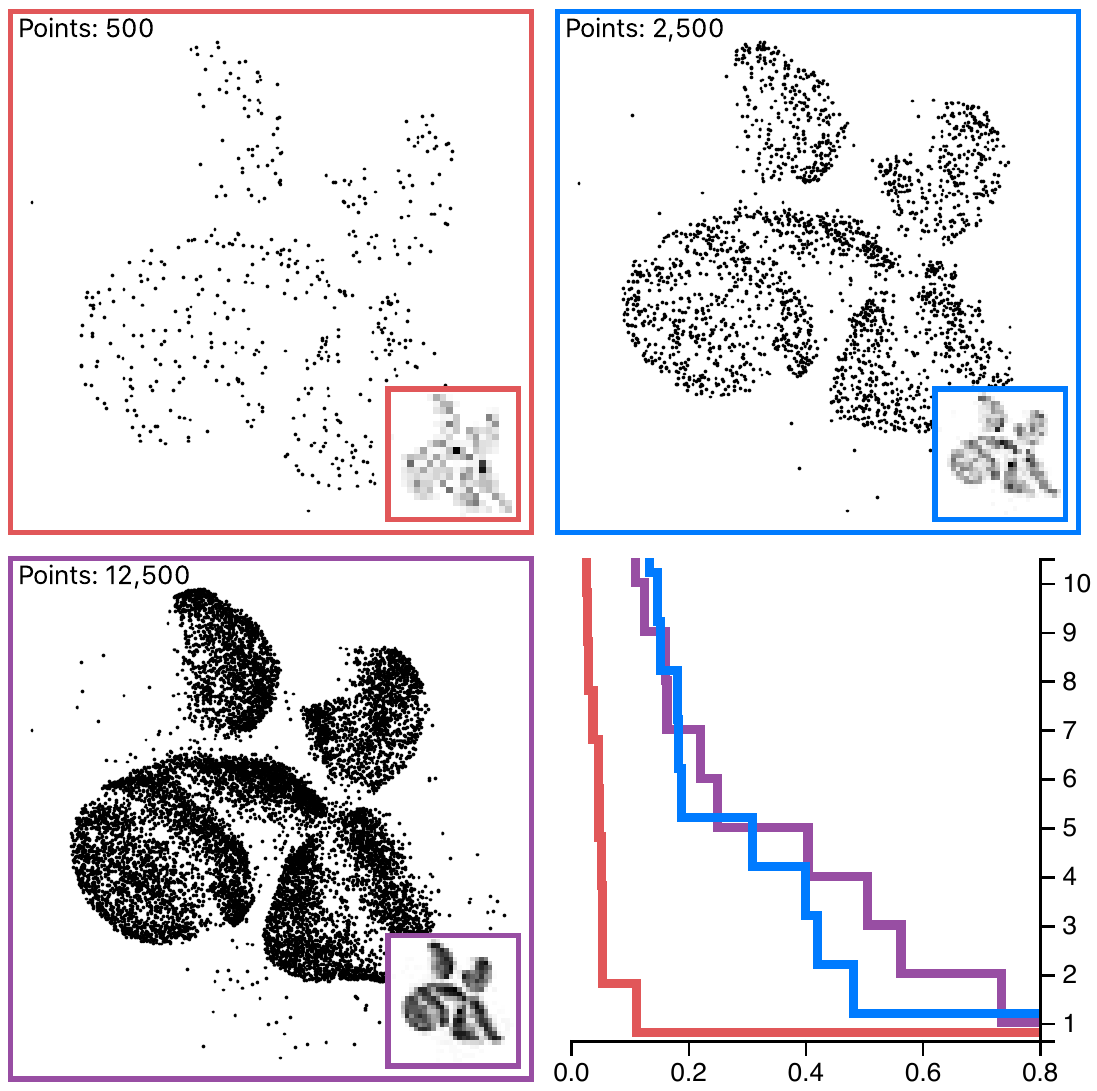}
    \hfill
    \includegraphics[height=179pt]{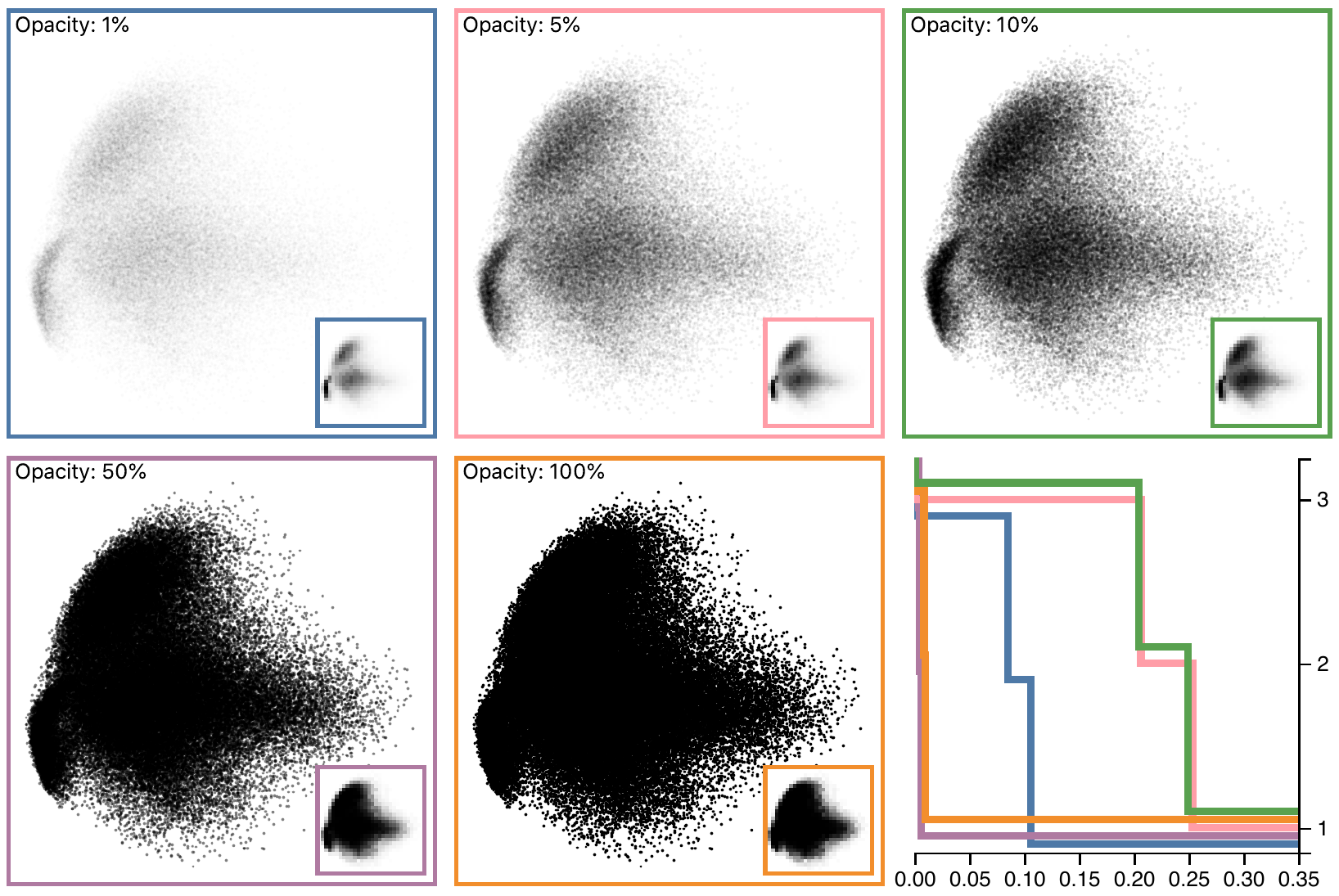}
    
    \vspace{-10pt}
    {\subfigure[Varying the Number of Points\label{fig:teaser:a}]{\hspace{0.36\linewidth}}}
    \hfill
    {\subfigure[Varying the Opacity of Points\label{fig:teaser:b}]{\hspace{0.56\linewidth}}}
  
    \caption{Demonstration of our density-based model being used to evaluate how differentiable clusters are in scatterplots that vary (a)~the number of points shown ($500$, $2500$, and $12500$) or (b)~the opacity of points ($1\%$, $5\%$, $10\%$, $50\%$, and $100\%$). The threshold plots (lower right) show how easy it is to visually identify (horizontally, bigger is better) a certain number of clusters (vertically). For example, when varying the opacity~(b), the threshold plot shows that 3 clusters are most clearly visible in the pink ($5\%$ opacity) and green ($10\%$ opacity) scatterplots, significantly less visible in the blue ($1\%$ opacity) scatterplot, and not visible at all in the purple ($50\%$ opacity) or orange ($100\%$ opacity) scatterplots. Designers can use these threshold plots to select the visual encodings that maximize the clarity of data.} 
	\label{fig:teaser}
}

\definecolor{red}{rgb}{0.6, 0, 0}
\definecolor{blue}{rgb}{0, 0, 0.6}
\definecolor{green}{rgb}{0.16, 0.435, 0.16}

\newcommand{\Dspace}        {{\mathbb D}}

\definecolor{lightergray}{gray}{0.925}

\newcommand{\para}[1]{\paragraph{#1}}

\makeatletter
\let\orgdescriptionlabel\descriptionlabel
\renewcommand*{\descriptionlabel}[1]{%
  \let\orglabel\label
  \let\label\@gobble
  \phantomsection
  \edef\@currentlabel{#1\unskip}%
  \let\label\orglabel
  \orgdescriptionlabel{#1}%
}
\makeatother

\vgtcinsertpkg

\begin{document}


\firstsection{Introduction}
\label{sec:intro}

\maketitle

Scatterplots are commonly used to reveal several types of relationships between quantitative variables~\cite{friendly2005early}. Numerous perceptual studies have evaluated the effectiveness of scatterplots in low-level tasks that include assessing trends~\cite{doherty2007perception, nguyen2016visual}, measuring correlation~\cite{rensink2010perception, harrison2014ranking, best2006perceiving}, and average and relative mean judgments~\cite{gleicher2013perception}. Clustering, in particular, is an aggregate-level task~\cite{munzner2014visualization, sarikaya2018scatterplots,liao2018cluster} that has been utilized in a variety of applications, e.g., weather forecasting, text analysis, and large-scale data analysis~\cite{szafir2016four, wong2014aligning, li2018concavecubes, ware2013designing}. Clustering occurs when patterns in the data form distinct groups~\cite{amar2005low,sarikaya2018design}. However, at its core, clustering is an ill-posed problem, as the ``correct'' clustering depends upon multiple factors~\cite{domany1999superparamagnetic,fern2011multiclust}.

When considering clustering in scatterplots, several factors play a role in how they are perceived. Data aspects, such as the {data distribution} size/type and the {number of data points}, can influence the visual presentation. On the other hand, visual encoding properties, such as {mark type}, {size}, and {opacity}, have the potential to influence perceptual judgments~\cite{cleveland1984graphical}. What is not well understood is how these various factors, \textit{many of which are under the control of the visualization designer}, influence the perception of clusters. The presentation of data is particularly important when considering that a biased representation of the data may provide an inaccurate summary, leading to invalid conclusions~\cite{szafir2018modeling, li2010size, gramazio2014relation, kim2018assessing}.

In this paper, we explore the multi-factor judgments used in identifying clusters in scatterplots through a crowdsourced user study. Based upon this study, we develop 2 models for the perception of clusters in scatterplots, using a data structure from Topological Data Analysis, called the \textit{merge tree}~\cite{wasserman2018topological}. We validate the models on a variety of variables---\textit{the number of points}, \textit{cluster distribution size}, \textit{size of data points}, and \textit{opacity of data points}---to verify the accuracy of the models and analyze their effects. Our results show that the perception of the number of clusters does indeed depend upon all 4 factors. Moreover, we show that the merge tree-based models do match an average user's perception of the clusters in a given scatterplot.

Finally, we demonstrate how the models can be used to optimize visualization designs. While some variables, such as distribution size, are difficult to control in a visualization, designers can use our models and findings as a guideline to balance the design factors that they do have control over---the number of points shown (e.g., via subsampling~\cite{hu2019data, chen2019recursive}), data point size~\cite{szafir2016four,urribarri2017prediction}, or opacity~\cite{matejka2015dynamic, micallef2017towards}---to optimize the saliency of the clusters in a visualization. A demo of the approach can be viewed at {\small \textless\textcolor{blue}{\url{https://usfdatavisualization.github.io/TopoClusterPerceptionDemo}}\textgreater}.

\section{Prior Work}
\label{sec-label:related}

We provide brief coverage of clustering in scatterplots and perception of the visual factors evaluated in our study.

\subsection{Clustering in Scatterplots}

    Clustering plays an important role in exploring and understanding many types of data~\cite{sarikaya2018scatterplots, sarikaya2018design}. A design factor survey defined clustering as a {high-level data characterization}---the ability to identify groups of similar items~\cite{sarikaya2018design}. Amar et al.\ presented a set of tasks for visual analytics that defined clusters as having ``similar attribute values in a given set of data''~\cite{amar2005low}. 

    \para{Taxonomies of Clustering Factors}
    Identifying clusters is directly influenced by the perception of cluster separation, and much of our understanding has come from studying dimension reduction (DR) techniques.  Lewis et al.\ compared the effectiveness of DR techniques using human judgments and concluded that T-SNE performs better than other commonly used methods when expecting clusters in the data \cite{lewis2012behavioral}.  Etemadpour et al.\ showed, however, the performance of DR techniques also depends on data characteristics~\cite{etemadpour2014perception}, e.g., the separability of clusters, and later created a user-centric taxonomy of visual tasks related to clustering in DR techniques~\cite{etemadpour2015user}.  A taxonomy of visual cluster separation in scatterplots used a qualitative evaluation to identify 4 important factors---scale, point distance, shape, and position~\cite{sedlmair2012taxonomy}. The taxonomy gives a context to our visual factor selection. Sedlmair and Aupetit later evaluated 15 class separation measures for assessing the quality of DR using human input for building a machine learning framework~\cite{sedlmair2015data} and later extended the framework to include an even greater number of measures~\cite{aupetit2016sepme}.

    \para{Perceptual Models of Clustering}
    Several works have considered how to model the perception of clusters.  For example, a recent study that used eye-tracking to analyze user perception in cluster identification, highlighted the role of Gestalt principles, especially proximity and closure~\cite{etemadpour2014eye}.  Matute et al.\ provided a method to quantify and represent scatterplots through skeleton-based descriptors that measured scatterplot similarity~\cite{matute2017skeleton}. However, their approach does not consider visual encodings in the evaluation.  ScatterNet, a deep learning model, captures perceptual similarities between scatterplots to emulate human clustering decisions but lacks explainability in the choices~\cite{ma2018scatternet}.  The scagnostics technique focused on identifying the patterns in scatterplots, including clusters~\cite{dang2014transforming}. However, a study by Pandey et al.\ showed that they do not reliably reproduce human judgments~\cite{pandey2016towards}.  Recently, ClustMe used visual quality measures to model human judgments to rank scatterplots~\cite{abbasclustme}. ClustMe performed well in reproducing human decisions for clustering patterns. In contrast, we are studying the extent to which various factors influence the perception of clusters and building explainable models of how humans perceive cluster separation using the merge tree data structure.

    \para{Clustering in Non-Scatterplot Contexts}
    Clustering has been studied in other types of visualization, such as text~\cite{alexander2014serendip}, maps~\cite{ware2013designing, li2018concavecubes}, and bubble charts~\cite{szafir2016four}.  A task-based evaluation found that on small data, bar and pie charts outperformed tables, scatterplots, and line charts in clustering tasks~\cite{saket2018task}.  The performance in cluster perception in pie charts is traced back to its effectiveness in facilitating proportional judgments through a part-whole relationship~\cite{eells1926relative,spence1991displaying}. Similarly, we hypothesize that the relative distance between clusters and the relative density of the image influence cluster identification.

\subsection{Factor Selection on Scatterplots}

    Several prior perceptual studies have demonstrated the effect of visual encodings on analysis tasks~\cite{szafir2018modeling, gramazio2014relation, cleveland1984graphical}. A variety of factors influence group or separation perception~\cite{wong2010points}, including color, size, shape~\cite{sedlmair2012taxonomy}, orientation~\cite{cohen2008perceptual}, texture~\cite{anobile2016number}, opacity \cite{micallef2017towards}, density~\cite{wilkinson2005graph}, motion and animation~\cite{etemadpour2017density,veras2019saliency,chen2018using}, chart size~\cite{heer2009sizing}, and others. Other studies have demonstrated a perceptual effect in scatterplots when changing factors in the data, including data distribution types, number of points, the proximity of concentrations of points, data point opacity, and relative density~\cite{gramazio2014relation, chung2016ordered, kim2018assessing, szafir2018modeling, gleicher2013perception, sadahiro1997cluster,correll2018looks}.  Overdraw in scatterplots, in particular, has been addressed with a variety of techniques, e.g., splatterplot~\cite{mayorga2013splatterplots}, recursive sampling~\cite{chen2019recursive}, set cover optimization~\cite{hu2019data}, feature-preserving visual abstraction~\cite{chen2014visual}, or by applying various clutter reduction techniques~\cite{ellis2007taxonomy}, e.g., sampling~\cite{wei2010multi, dix2002chance, ellis2002density, ellis2005sampling} or changing opacity~\cite{matejka2015dynamic}.

    From this collection of possible factors, \textit{we focus our study specifically on the factors that most influence visual density}, including the distribution of and distance between concentrations of points, the number and size of data points, and data point opacity in the visualization.

    \para{Point Distribution} 
    Several prior studies have investigated the influence of the distribution of data points on cluster perception. An early study of 8 participants on 24 homogeneous dot patterns studied the impact of varying densities and gaps between 2 square-shaped clusters~\cite{o1974human}. Sadahiro later developed a mathematical model to represent cluster perception in point distributions based on 3 factors---proximity, concentration, and density change---and suggested perception is significantly influenced by the concentration and density change~\cite{sadahiro1997cluster}. Similarly, the scagnostics density property identifies concentrations of points directly influenced by the distribution of points~\cite{wilkinson2005graph}.

    \para{Number of Data Points} 
    Sadahiro also showed that the higher the number of points in a given area, the higher the chances are that they would be perceived as a cluster, due to increased density~\cite{sadahiro1997cluster}. Gleicher et al.'s empirical study asked participants to compare and identify average values in multi-class scatterplots~\cite{gleicher2013perception}. It demonstrated that judgments are improved with a higher number of points.

    \para{Size of Data Points}
    The size of symbols is an important factor in visual aggregation tasks in scatterplots~\cite{szafir2016four}. As the size of data points increases, so does the density, which directly influences cluster perception~\cite{sadahiro1997cluster}.  Symbol size also has a direct influence on discriminability in certain tasks~\cite{li2010size}, e.g., in color perception tasks~\cite{stone2012visualization}. Szafir's study on color-difference perception found that perceived color difference varies by the size of marks~\cite{szafir2018modeling}. Size also influences search task effectiveness. Gramazio et al.'s study on target search demonstrated that while the quantity of data points has little effect on searching for a target, increasing symbol size reduces search time in a display of random points~\cite{gramazio2014relation}.

    \para{Opacity of Data Points}
    As the number of data points increases, scatterplots suffer from overplotting, which obscures the data distribution. Reducing mark opacity can alleviate overplotting to aid various visual analytics tasks~\cite{sarikaya2018scatterplots}, e.g., spike detection in dot plots~\cite{correll2018looks}.  Furthermore, different opacity levels aid in different visual tasks---while low opacity benefits density estimation for large data, it also makes locating outliers more difficult~\cite{micallef2017towards}. Matejka et al.\ defined an opacity scaling model for scatterplots that is based on the data distribution and crowdsourced responses to opacity scaling tasks~\cite{matejka2015dynamic}. Still, their study did not evaluate how a scatterplot design based on data symbol opacity can affect user performance on visual analysis tasks. Somewhat related to opacity is luminance, which can be modeled using extreme end lightness~\cite{li2010model}, creating a popout effect~\cite{gutwin2017peripheral}.

\section{Study Methodology}
\label{sec-label:methodology}

We investigate how visual factors affect subject responses in the task of counting the number of clusters in a scatterplot. From this, we build and analyze two models to estimate the number of clusters an average user would perceive. One model is based on the separation distance between distributions, and the other uses the visual density of points.

\subsection{Factors}
\label{sec-label:factors}

Data are presented as point marks (i.e., circles 
\begin{tikzpicture}
    \draw [black,fill=black](1,1) circle (0.1cm);
\end{tikzpicture}) 
on the scatterplots and groups of similar objects form {clusters}. Based on our review of prior work, we chose to use a normal distribution to generate clusters, and we selected the following experimental factors:

\noindent
\begin{minipage}{0.5\linewidth}
    \vspace{5pt}
    \noindent\hspace{10pt}1.~\underline{Distribution size} ($S$)
 
    \vspace{10pt}
    \noindent\hspace{10pt}2.~\underline{Number} of data points~($N$)
 
    \vspace{10pt}
    \noindent\hspace{10pt}3.~\underline{Size} of data points~($P$)
 
    \vspace{10pt}
    \noindent\hspace{10pt}4.~Data point \underline{opacity} ($O$)
 
\end{minipage}
\hspace{6pt}
\begin{minipage}{0.45\linewidth}
    \includegraphics[width=\linewidth]{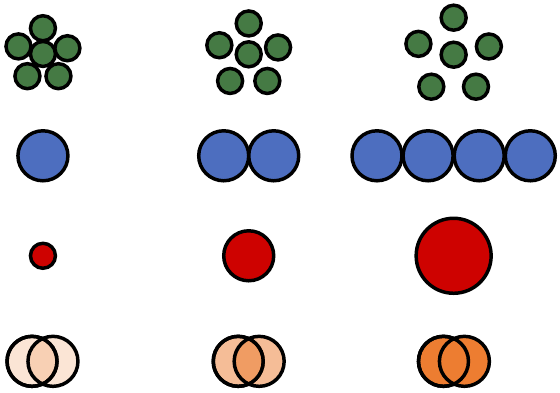}
\end{minipage}

\subsection{Experiments Setup}

    We designed our experimental study in 3 stages: 
        (1)~a preliminary experiment to calibrate the experimental factors;
        (2)~a crowdsourced Amazon's Mechanical Turk (AMT) experiment to validate our models; and 
        (3)~a follow-up AMT study to elaborate upon 1 of our models.

\subsection{Data Generation}
\label{sec-label:methodology:data}

\begin{figure}[!h]
    \centering
    \begin{minipage}[t]{0.425\linewidth}
        \hfill
        \includegraphics[width=0.9\linewidth]{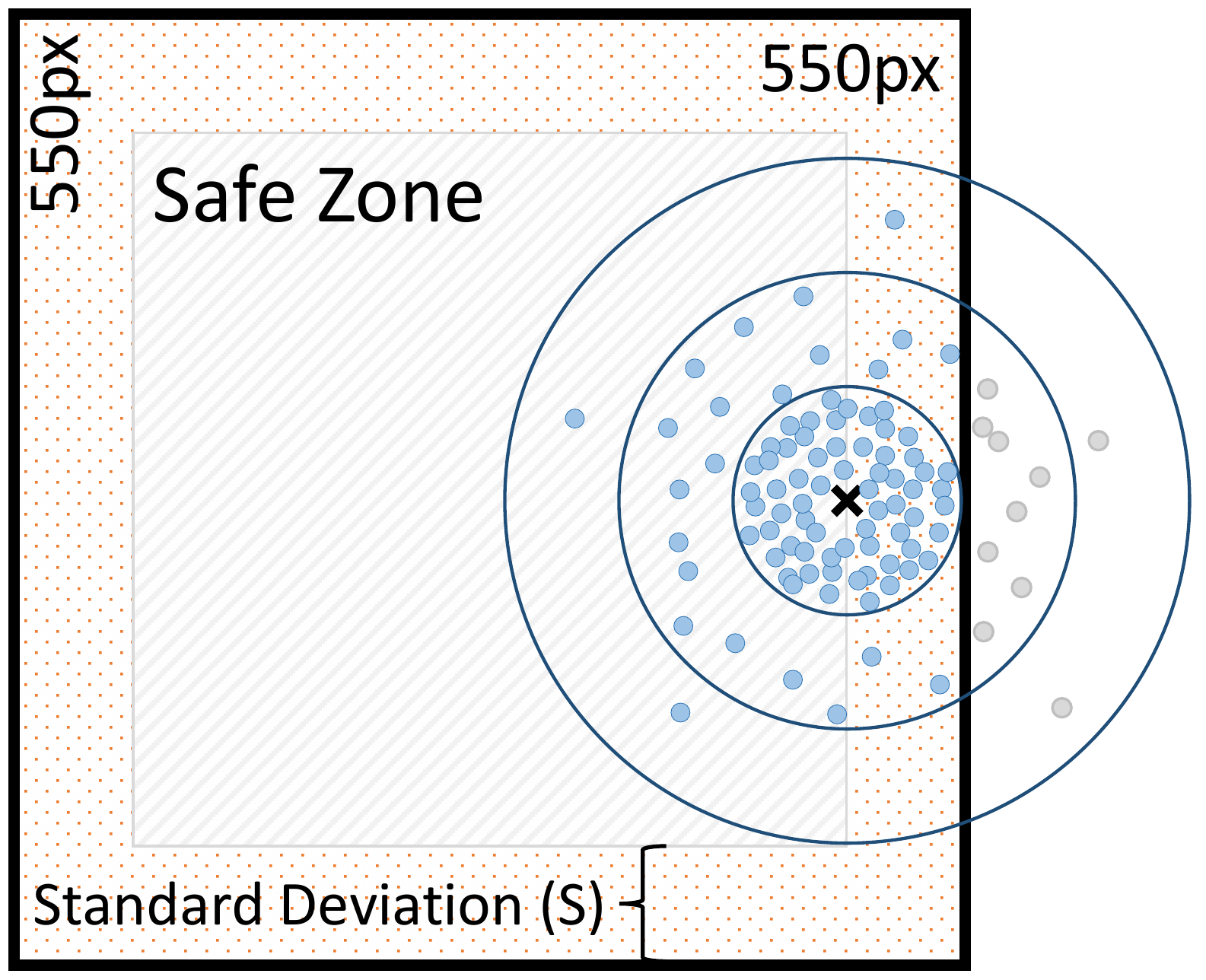}
    \end{minipage}
    \hfill
    \begin{minipage}[t]{0.52\linewidth}
        \vspace{-86pt}
        \caption{Illustration of the data generation. All plots are $550\times550$. Cluster centers are placed within a ``safe zone'', 1 standard deviation away from the boundary. Points are sampled from a normal distribution (blue), and points outside of the image are discarded without replacement (gray).}
        \label{fig:data-generation}
    \end{minipage}
\end{figure}

    Datasets are synthesized using 5 input parameters (see \autoref{fig:data-generation}): stimuli dimensions~($[X \times Y]$~pixels); number of clusters~($C$); distribution size, i.e., standard deviation~($S$~pixels); number of points~($N$); and signal-to-noise ratio~($SNR$). First, $C$ cluster centers are randomly placed within a ``safe zone'' defined as 1 standard deviation from the stimuli (image) border, in other words, $x\in[S,X-S]$ and $y\in[S,Y-S]$. Each cluster is assigned an equal share of the available points ($N/C$). Points are randomly placed around their cluster center using a {normal distribution} with a standard deviation of $S$ pixels. Points outside of the image dimensions are discarded without replacement. Next, an additional $N/SNR$ points representing noise are placed randomly using a uniform distribution across the image dimensions. Finally, to generate images, 2 more input parameters are used: point size~($P$~pixels) and point opacity~($O$). The points are drawn as filled circles of $P$ area with $O$~opacity. Example stimuli are shown in \autoref{fig:example_stimuli}.

    \vspace{5pt}\noindent 
    In all experiments some inputs were kept constant:
    \begin{itemize}[noitemsep,itemsep=4pt]
 
        \item \uline{Stimuli dimensions ($[X \times Y]$)}: $[550_{px} \times 550_{px}]$ --- The vertical size was selected such that the image would fit on the majority of desktop monitors without scrolling~\cite{StatCounter2019}. The horizontal resolution was selected to match, avoiding any directional bias.
 
        \item \uline{Signal-to-noise ratio ($SNR$)}: $10:1$ --- We manually optimized the $SNR$ by looking for a high level of noise that would not overwhelm the clusters. We ended at $10:1$, making the maximum total number of data points in any given dataset $N+0.1 \cdot N$. 
 
    \end{itemize}

\newcommand{\placeImageLabelC}[3]{\put(-60,18){
    \transparent{0.8}
 		\colorbox{white}{\begin{minipage}[t][15pt][t]{16pt}\mbox{}
 		\end{minipage}}
 		}\put(-57,15){
 		\begin{minipage}[t][0pt][t]{0pt}
 			\tiny 
 			\mbox{#1} \\
 			\mbox{#2} \\
 			\mbox{#3}
 		\end{minipage}
 		}}

\begin{figure}[!t]
    \centering
    \begin{tabular}{@{}c@{\hspace{2pt}}c@{\hspace{2pt}}c@{\hspace{4pt}}|@{\hspace{4pt}}c@{}}
        \begin{minipage}[t]{0.235\linewidth}\centering
     	    {\includegraphics[trim=55pt 55pt 55pt 55pt, clip, width=\linewidth]{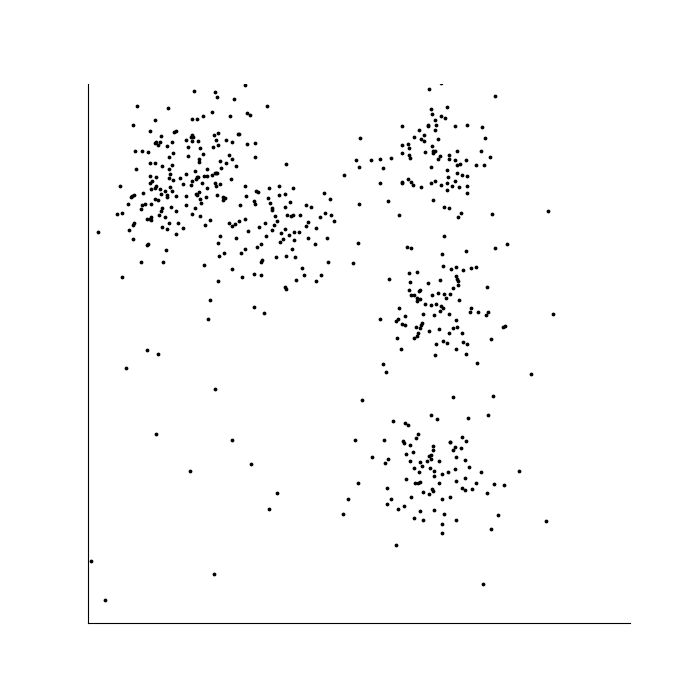}}
     	    \placeImageLabelC{$S$$=$$40_{px}$}{$N$$=$$500$}{$P$$=$$3_{px}$}
     	\end{minipage}
     	&
        \begin{minipage}[t]{0.235\linewidth}\centering
     	    {\includegraphics[trim=55pt 55pt 55pt 55pt, clip, width=\linewidth]{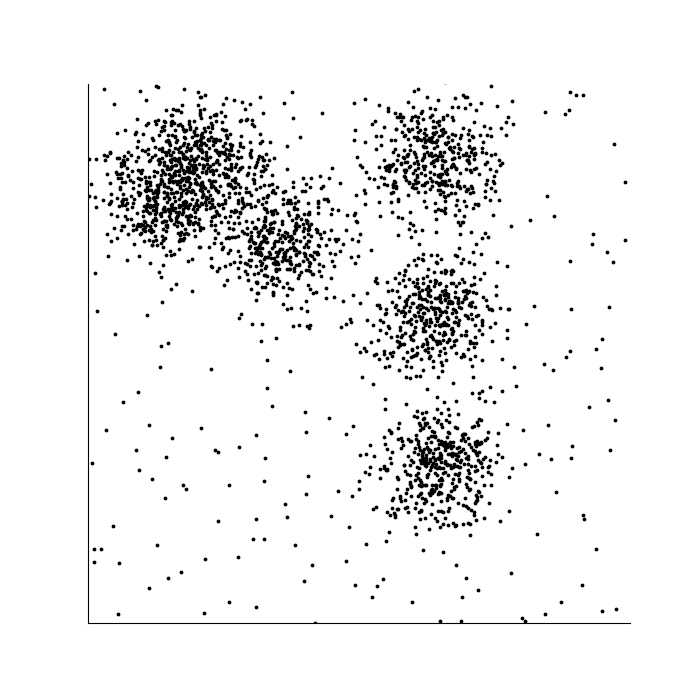}}
     	    \placeImageLabelC{$S$$=$$40_{px}$}{$N$$=$$2500$}{$P$$=$$3_{px}$}
        \end{minipage}
     	&
        \begin{minipage}[t]{0.235\linewidth}\centering
     	    {\includegraphics[trim=55pt 55pt 55pt 55pt, clip, width=\linewidth]{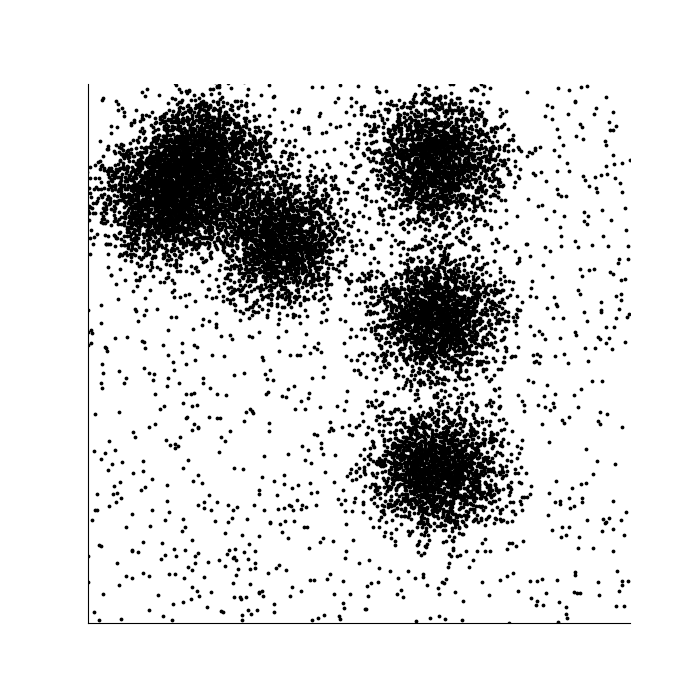}}
     	    \placeImageLabelC{$S$$=$$40_{px}$}{$N$$=$$12500$}{$P$$=$$3_{px}$}
     	\end{minipage}
     	&
        \begin{minipage}[t]{0.235\linewidth}\centering
     	    {\includegraphics[trim=55pt 55pt 55pt 55pt, clip, width=\linewidth]{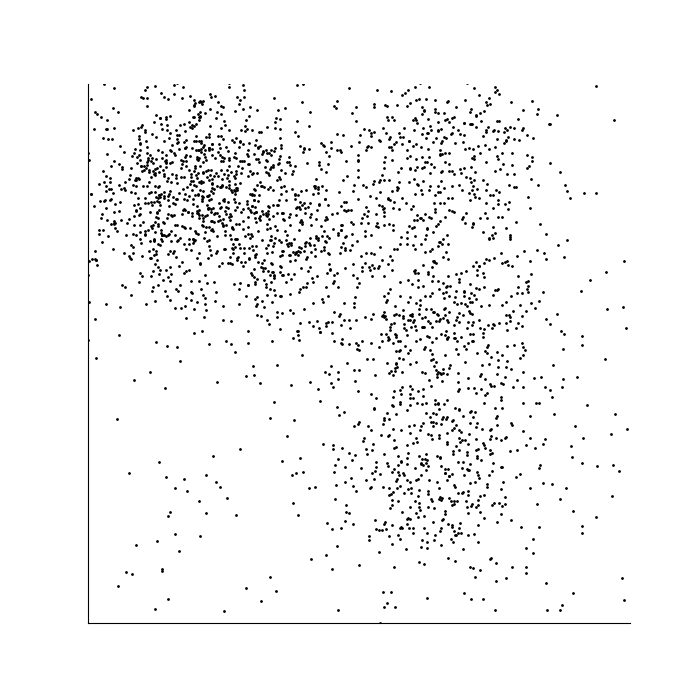}}
     	    \placeImageLabelC{$S$$=$$70_{px}$}{$N$$=$$2500$}{$P$$=$$1_{px}$}
     	\end{minipage} 	
     	\\
     
        \begin{minipage}[t]{0.235\linewidth}\centering
     	    {\includegraphics[trim=55pt 55pt 55pt 55pt, clip, width=\linewidth]{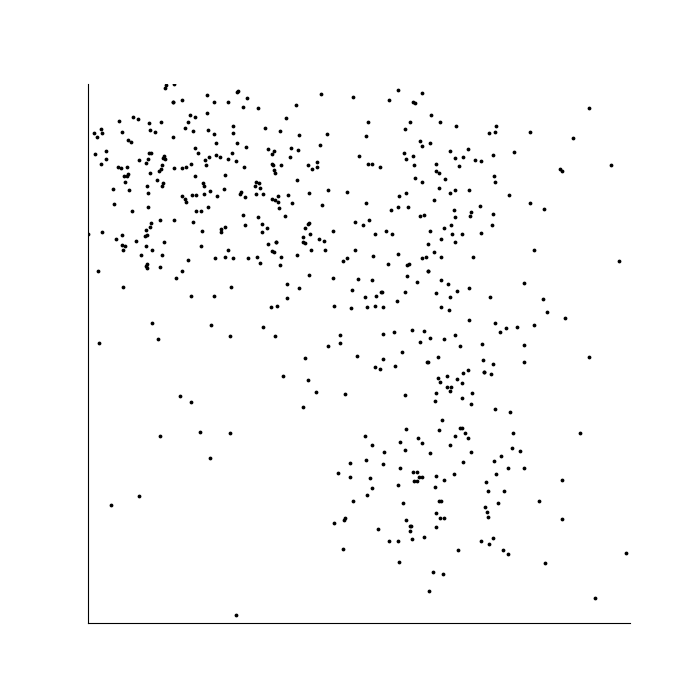}}
     	    \placeImageLabelC{$S$$=$$70_{px}$}{$N$$=$$500$}{$P$$=$$3_{px}$}
     	\end{minipage}
     	&
        \begin{minipage}[t]{0.235\linewidth}\centering
     	    {\includegraphics[trim=55pt 55pt 55pt 55pt, clip, width=\linewidth]{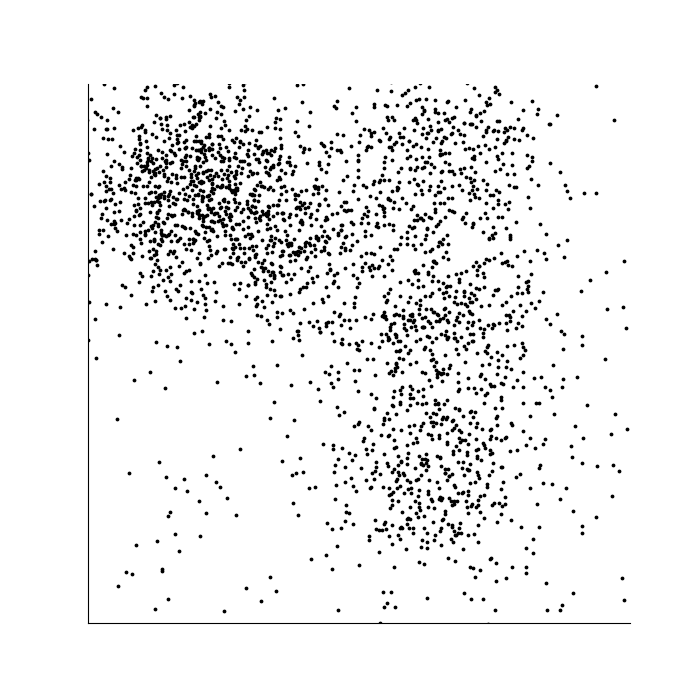}}
     	    \placeImageLabelC{$S$$=$$70_{px}$}{$N$$=$$2500$}{$P$$=$$3_{px}$}
     	\end{minipage}
     	&
        \begin{minipage}[t]{0.235\linewidth}\centering
     	    {\includegraphics[trim=55pt 55pt 55pt 55pt, clip, width=\linewidth]{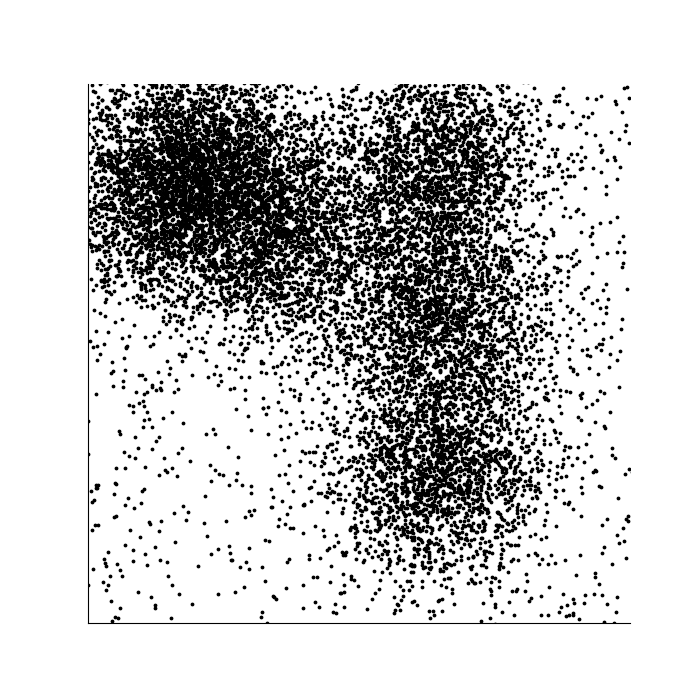}}
         	\placeImageLabelC{$S$$=$$70_{px}$}{$N$$=$$12500$}{$P$$=$$3_{px}$}
     	\end{minipage}
     	&
     	\begin{minipage}[t]{0.235\linewidth}\centering
     	    {\includegraphics[trim=55pt 55pt 55pt 55pt, clip, width=\linewidth]{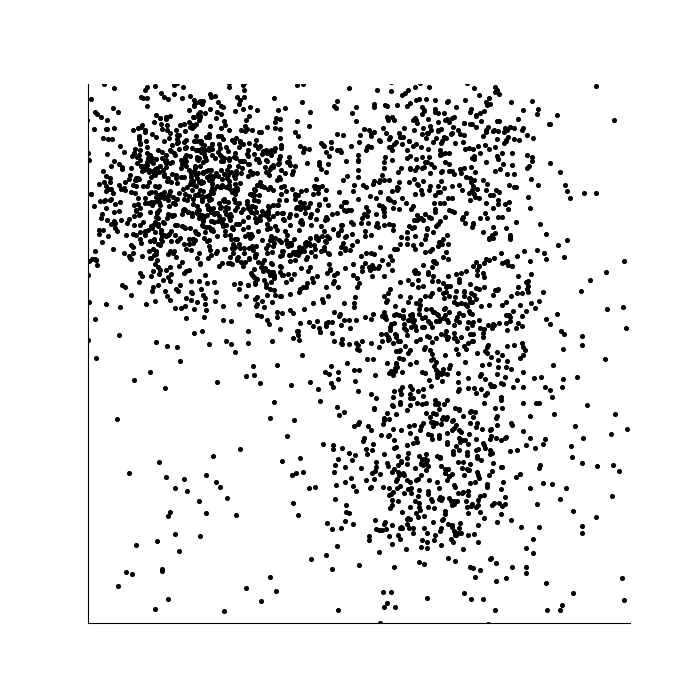}}
     	    \placeImageLabelC{$S$$=$$70_{px}$}{$N$$=$$2500$}{$P$$=$$7_{px}$}
     	\end{minipage}
     	\\
    \end{tabular}
 
    \caption{Example stimuli with the same cluster centers ($C=5$), but varying distribution size ($S$), number of points ($N$), and size of points ($P$).} 
 	\label{fig:example_stimuli}
 
 \end{figure}

\section{Preliminary Experiment}
\label{sec-label:preliminary}

We performed a preliminary user study to test initial hypotheses and calibrate parameters for the larger AMT experiment. Based on our observation and study of prior work, we drafted the following hypotheses: 
\begin{description}[leftmargin=!,labelindent=5pt,itemindent=-20pt]

    \item[{[H1]}\label{Hyp:H1}] The distribution size of clusters affects the accuracy in cluster count identification in scatterplots.
    
    \item[{[H2]}\label{Hyp:H2}] The number of data points affects the accuracy in cluster count identification in scatterplots.
    
\end{description}

\subsection{Properties and Data Generation}
\label{sec-methodology-factors}

As aligned with previous empirical studies, e.g.,~\cite{saket2018task}, we selected parameter values to maintain a reasonable level of difficulty. We designed the task such that the response time for a single stimulus would be 5~to 20~seconds. We selected the following experimental factors:
\begin{itemize}[noitemsep,itemsep=4pt]
    \item \uline{Number of clusters ($C$)}: $\{4-12\}$ --- The number of clusters was selected using trial-and-error to avoid tasks that were too easy (i.e., trivial to count) or too difficult (i.e., larger number or sparse clusters).
    
    \item \uline{Data point size/area ($P$)}: $\{20_{px}\}$ --- Experimental calibration was not needed for point size, as reasonable values could be determined analytically. Therefore, the point size was fixed in order to calibrate other factors. 
    
    \item \uline{Number of data points ($N$)}: $\{1000,5000,10000\}$
    
    \item \uline{Distribution size ($S$)}: $\{20_{px}, 35_{px}, 50_{px}, 65_{px}, 90_{px}\}$ --- $N$ and $S$ were the main factors to test/calibrate. The value ranges were selected using our observation of sample stimuli and judgment of factors from prior work, considering sufficient range, minimum and maximum values, and the number of experimental conditions that could be reasonably tested. 
    
    \item \uline{Data point opacity ($O$)}: $\{100\%\}$ --- Points were fully opaque.
    
\end{itemize}

\noindent The dependent variable we tested was:
\begin{itemize}
    \item \uline{User-selected number of clusters ($U$)}: $[1-15]$
\end{itemize}

Dataset generation for the preliminary experiment was done in the following manner---for every combination of $S$ and $N$, $500$ stimuli (i.e., images) were generated with a random number of clusters, $C$. Other parameters were fixed as described, leading to a pool of $|S| \times |N| \times 500 = 7500$ stimuli.

\subsection{Study Procedure}
\label{sec-label:preliminary:proc}

We developed a webpage for the experiments, where each participant was shown 50 images from the pool of $7500$, one at a time, and asked the number of clusters they could see. Answers were recorded using a drop-down box with options $1-15$. The maximum allocated time for each task was 20 seconds. At the expiration of time, the page was automatically advanced. To mitigate any effects or bias, we placed a blank screen between every 2 tasks~\cite{healey2012attention}. At the beginning of the experiment, we included a brief introduction to clustering and 3 training tasks for each participant, which were similar to the study tasks that followed. The experiment was expected to last 20-30~minutes, including demographic details and training tasks.

We recruited 30 participants from the College of Engineering at the University of South Florida for the IRB approved study. Participant ages ranged from 18-28 ($\mu_{age}$=23), with 24~males and 6 females. No compensation was provided. In total, $50$ trials $\times$ $30$ participants = $1500$ responses were collected. While performing data quality checks on the responses, we found discrepancies---participants responding to stimuli in less than 1 second or those with responses of 1 cluster to all stimuli---in 4 participants results and removed them from analysis, leaving $1300$ responses ($26 \times 50)$. We further identified and removed $161$  stimuli that had been reused from the pool only keeping the first occurrence\footnote{We acknowledged this is a flaw in our preliminary study design. However, since our primary goal was parameter calibration, the experiment still has value. We avoid this bias in our AMT experiment by generating stimuli per participant.}, leaving  $1139$ responses for analysis.

\subsection{Analysis and Result}

\begin{figure}[!b]
    \centering
    
    \begin{minipage}[t]{0.0325\linewidth}
        \rotatebox{90}{
		    \begin{minipage}{2.35cm}
		    \centering
		    \footnotesize Frequency
		    \end{minipage}
		}
	\end{minipage}
	\hfill
	\begin{minipage}[t]{0.95\linewidth}	
	    \centering
        \includegraphics[width=\linewidth]{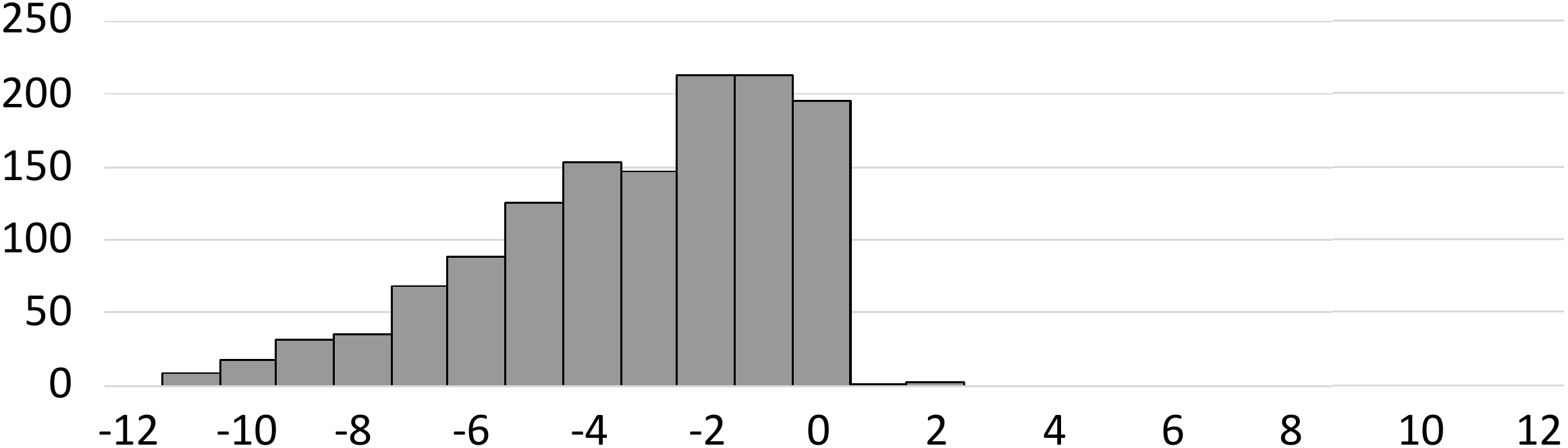}
        \footnotesize Differential ($\Dspace$)
	\end{minipage}

    \caption{The histogram of user responses \textit{differential}, $\Dspace$, against frequency for the preliminary experiment, appears as a truncated normal distribution.}
    \label{fig:prelim-histogram-womodel}
\end{figure}

\begin{figure*}[!tb]
    \centering
        
    \begin{minipage}[b]{0.335\textwidth}    
        \includegraphics[height=1.6in]{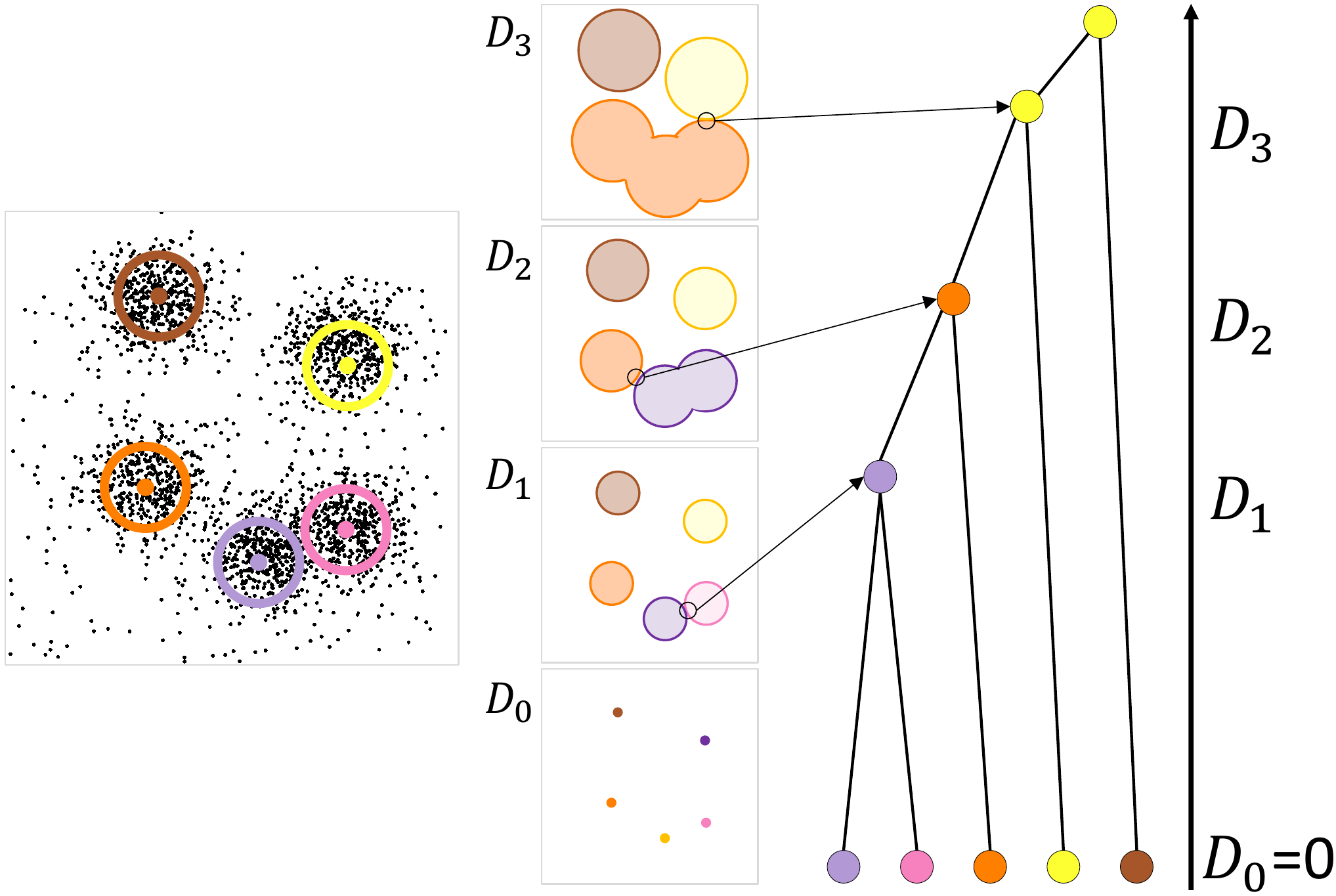}

        \vspace{-10pt}
        \hspace{30pt}
        \subfigure[\label{fig:model:ph:img}]{\hspace{10pt}}
        \hspace{35pt}
        \subfigure[\label{fig:model:ph:b}]{\hspace{10pt}}
        \hspace{30pt}
        \subfigure[\label{fig:model:ph:c}]{\hspace{10pt}}

        \vspace{-10pt}
        \caption{With the distance-based model, (a)~given the input data, (b)~cluster centers are analyzed~at a series of ball diameters. Connected components receive a unique color at each diameter. (c)~The scan of diameters leads to the creation of the merge tree, which marks topological events (i.e., creation and merging of components) with nodes.}
        \label{fig:model:ph}    
    \end{minipage}
    \hfill
    \begin{minipage}[b]{0.32\textwidth}    
        \includegraphics[height=1.60in]{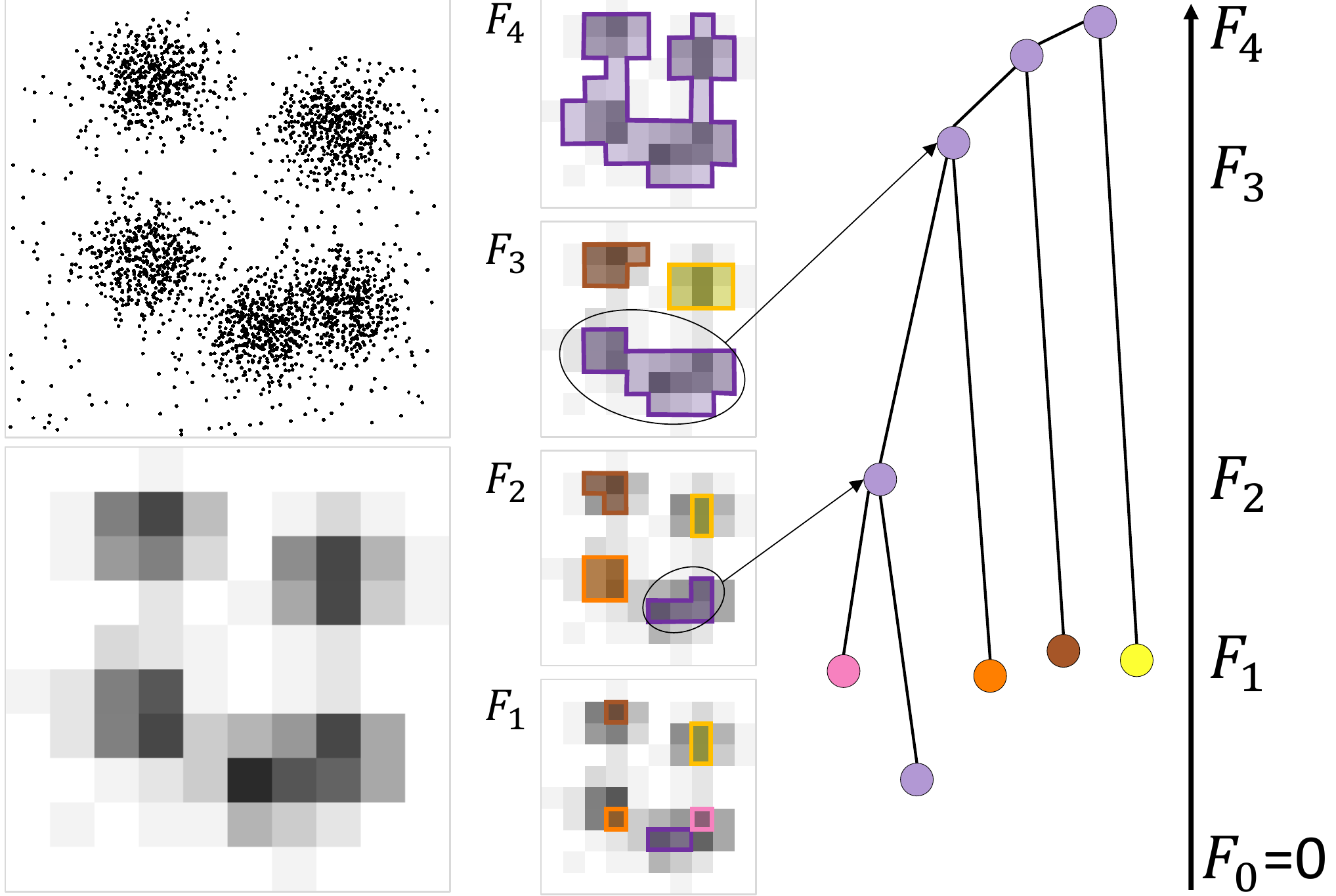}

        \vspace{-10pt}
        \hspace{30pt}
        \subfigure[\label{fig:model:ct:a}]{\hspace{10pt}}
        \hspace{35pt}
        \subfigure[\label{fig:model:ct:b}]{\hspace{10pt}}
        \hspace{30pt}
        \subfigure[\label{fig:model:ct:c}]{\hspace{10pt}}

        \vspace{-8pt}
        \caption{With the density-based model, (a)~given the input data (top), a density histogram (bottom) is calculated. (b)~The space is analyzed at different density values, and components are extracted. (c)~Tracking components across density values leads to the creation of the merge tree, which marks topological events with nodes.}
        \label{fig:model:ct}
    \end{minipage}
    \hfill
    \begin{minipage}[b]{0.312\textwidth}
        \centering

        \subfigure[Distance-based Model\label{fig:barcode:dist}]{\hspace{5pt}\includegraphics[width=0.54\linewidth]{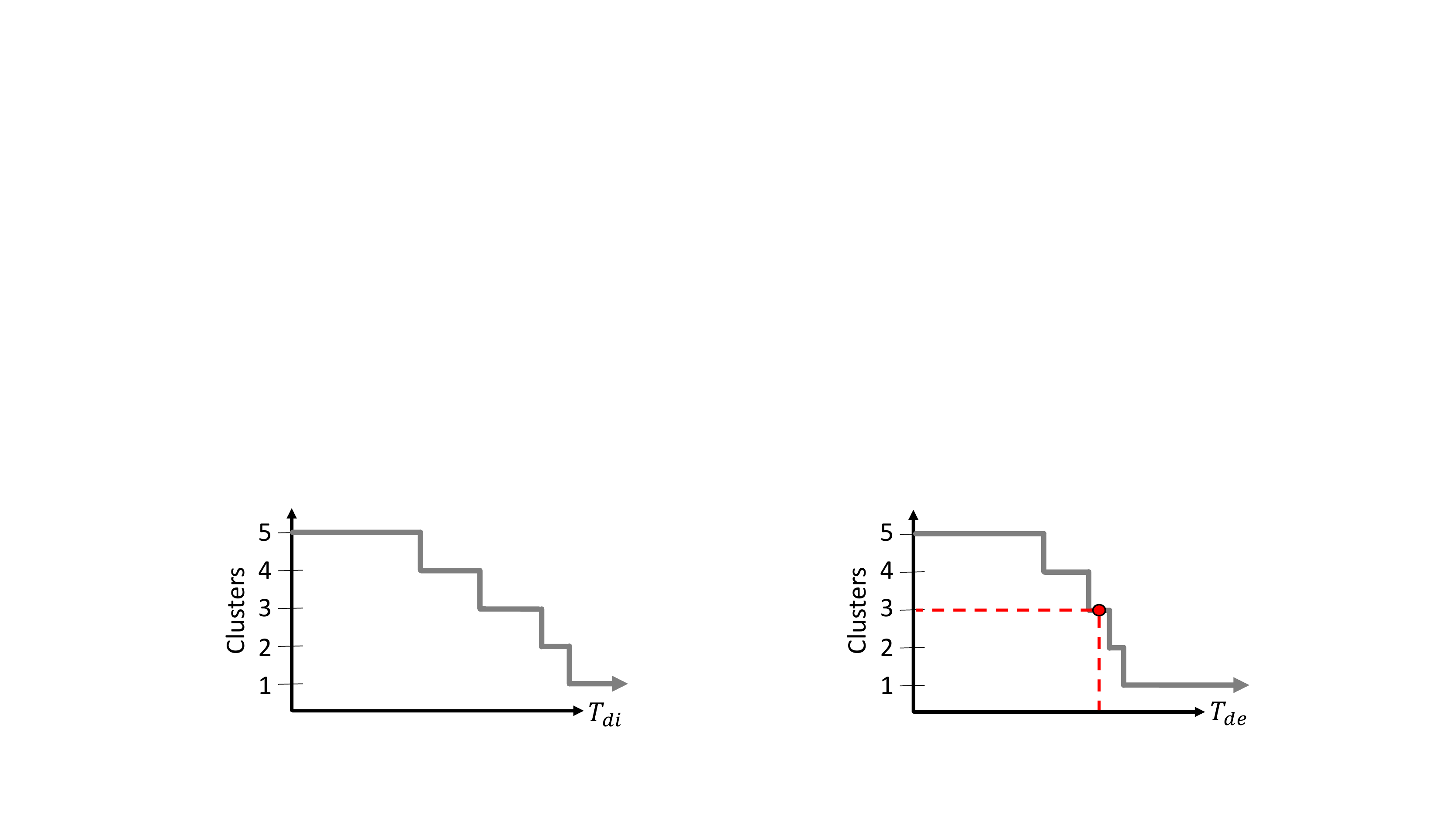}\hspace{5pt}}
        
        \subfigure[Density-based Model\label{fig:barcode:dens}]{\hspace{5pt}\includegraphics[width=0.54\linewidth]{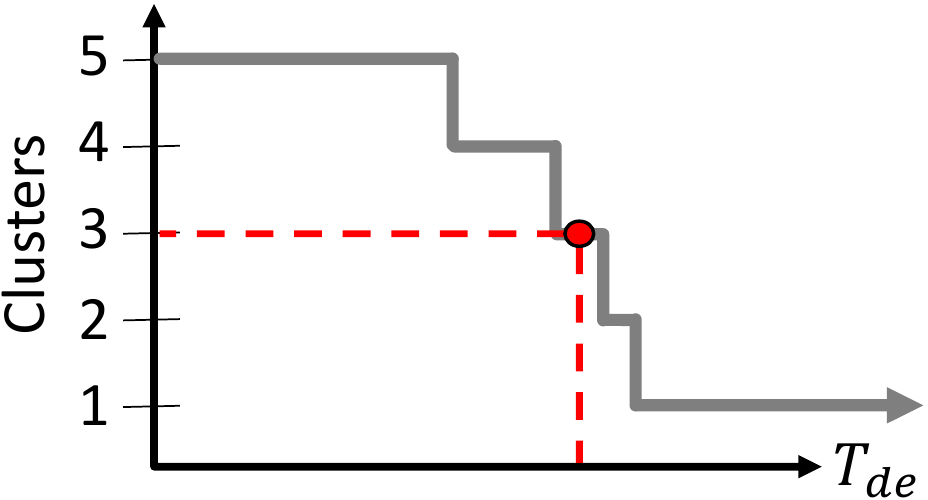}\hspace{5pt}}

        \caption{The persistence threshold plots for the (a)~distance- and (b)~density-based models in \autoref{fig:model:ph} and \autoref{fig:model:ct}, respectively. The horizontal axis represents the threshold, while the vertical axis shows the number of clusters.
        The red line shows how a threshold can be extracted from a given number of clusters and vice versa.}
        \label{fig:barcode}
    \end{minipage}
    
\end{figure*}

To measure accuracy for a given scatterplot,~$\tau$, we use the \textit{differential}: $\Dspace(\tau) = U_\tau - C_\tau$, where $U_\tau$ is the user response and $C_\tau$ is the number of clusters in the data. We analyzed the \textit{differential} against the independent factors \textit{distribution size}~($S$) and the \textit{number of data points}~($N$) using a 2-way ANOVA test. We also calculated partial eta-squared ($\eta^2$).
We observed that $S$ and $N$ have a significant effect on the accuracy in identifying the number of clusters, ($F_S(4,1130)=48.57, p< 0.01, \eta^2=0.12$) and  ($F_N(2,1130)=8.29, p< 0.01, \eta^2=0.02$), respectively. \textbf{These results confirm \ref{Hyp:H1} and \ref{Hyp:H2}}.

Although we found a significant effect, user accuracy had a low average $\mu_{\Dspace}=-3.27$ and a high standard deviation $\sigma_{\Dspace}=2.66$ (accurate predictions would have an average of $0$ with a small standard deviation). \autoref{fig:prelim-histogram-womodel} shows the histogram of differentials, which appear as a truncated normal distribution. The negative shift in the $\mu_\Dspace$ revealed that of the \textit{number of points} or \textit{distribution size} alone is insufficient to model the number of clusters users would perceive---an accurate model needs to consider the overlap of clusters. For example, in \autoref{fig:example_stimuli}, all images have an identical number of generated clusters, but the interaction between clusters causes differing numbers of clusters to appear. \uline{Instead, the distance between clusters or the visual density of the data needs to be considered as an additional factor in cluster perception modeling.} The next section introduces models that each considers one of these factors.

\section{Topology-based Modeling of Clustering}
\label{sec-label:topology}

We propose 2 models for capturing human perception of clusters based upon approaches from Topological Data Analysis (TDA)~\cite{wasserman2018topological}. TDA is a set of approaches used to study the ``shape'' of data, including scalar fields~\cite{rosen2019using,tierny2017topological}, vector fields~\cite{wang2013visualizing}, and high-dimensional data~\cite{choudhury2012topological,maljovec2016rethinking}.

Both models capture the clustering structure using a data structure called the \textit{merge tree}. The merge tree encodes a series of topological events in the form of creation and merging of components (specifically, $0$-dimensional homology groups), based upon properties of the space under a real-valued function. The first model, based upon the distance between cluster centers, is captured using a technique called \textit{persistent homology}~\cite{edelsbrunner2008persistent}. The second model, based upon the visual density of points, is captured by calculating the join tree of a scalar field~\cite{CarrSnoeyinkAxen2003}.

\subsection{Distance-based Model}

The distance-based model tries to capture human perception of clusters by considering the spatial resolution at which 2 or more cluster distributions will blend to be perceived as 1. We do this using the technique of persistent homology (PH)~\cite{edelsbrunner2008persistent}. We provide a simplified view of PH under our limited context. For a detailed introduction, see~\cite{EdelsbrunnerHarer2010}.

Construction begins with a finite set of points $V$ representing cluster centers embedded in Euclidean space (i.e., their positions on the scatterplot). Given a real number $D >= 0$, we consider a set of balls of diameter $D$ centered at points in $V$. Continuously increasing the diameter, $0 = D_0 \leq D_1 \leq D_2 \leq \cdots \leq D_m = \infty$, forms a $1$-parameter family of nested unions of balls. If at a given diameter $D_i$, 2 balls overlap, we consider these balls as a single component. \autoref{fig:model:ph:b} shows an example dataset with 4 values of $D_i$. As $D_i$ increases, more balls intersect and merge into larger components. At $D_\infty$, all balls will overlap, forming a single component.

To compute the PH, the points $V$ form the vertices of a graph. A $1$-simplex (an edge) is formed between 2 points in $V$ if and only if their balls intersect (i.e., the distance between them is $\leq D_i$). Sweeping $D_i$ from $0 \rightarrow \infty$, as $D_i$ increases, new edges are added to the graph. Components are efficiently calculated at each step by finding connected components of the graph using the set union data structure. The total computation time is $O(|E| \alpha(|V|))$, where $E$ are the edges of the graph, and $\alpha$ refers to the inverse Ackermann function, an extremely slow-growing function. At $D_\infty$, PH forms the complete graph. Therefore, there are $O(|V|^2)$ edges.

Creating the merge tree from the prior construction is relatively simple. The merge tree is parameterized with respect to $D$. At $D_0=0$, all cluster center components are \textit{born}. In other words, the balls have $0$ volume. These birth events appear in the merge tree as $1$ node per cluster, e.g., see the bottom of \autoref{fig:model:ph:c}. As $D_i$ increases, when 2 components first merge at a given $D_i$, a \textit{merge node} is added to the merge tree at $D_i$ connecting those components. For example, at $D_1$, the purple and pink components intersect, causing them to merge into a single component. From that point forward, 1 of the merged components \textit{dies} (i.e., no longer exists), while the other takes on the identity of the new merged component (in this context, it does not matter which). Referring back to \autoref{fig:model:ph:c}, when purple and pink merge at $D_1$, pink dies, while purple takes on the identity of the merged component. When the components finally merge into a single component, yellow in our example, this component dies at $\infty$. In other words, no matter how large the balls get, that 1 component will exist. It is also relevant to note that this particular construction has many parallels to single-linkage hierarchical clustering.

This model has 2 main limitations: (1)~it assumes that clusters are isotropic and have similar distributions; and (2)~it requires knowledge about the location of cluster centers. Our next model uses a related framework to overcome these limitations.

\subsection{Density-based Model}

The density-based model attempts to directly identify the \textit{relative} visual density at which users will differentiate between clusters. The density-based model is found by calculating the join tree of a scalar field. We again provide a simplified treatment---for a detailed description, see~\cite{CarrSnoeyinkAxen2003}.

First, a 2D histogram of the visual density is created for the scatterplot (i.e., a density plot). The image plane is divided into a set of grid cells of uniform width and height (selection of this resolution is discussed in our evaluation). Within each grid cell, the number of \textit{white} pixels is counted, and this is considered the density\footnote{We acknowledge this is not the usual calculation of density, e.g., see~\cite{dang2014transforming}, which would count the number of black pixels. However, our configuration makes the remainder of the discussion easier.}, $f_{xy}$. For illustrative purposes, this value is mapped to the range $F \in [0,255]$, where $0$ is empty (i.e., completely black) and $255$ is full (i.e., completely white), as shown in \autoref{fig:model:ct:a}.

The components of the density histogram are identified by sweeping $F$, such that $0=F_0 < F_1 < F_2 < \cdots < F_m =\infty$. At each $F_i$, histogram cells where $f_{xy}\leq F_i$ are extracted and components found by joining neighboring cells (we use the 8 surrounding neighbors). This is computed by treating histogram cells as graph nodes, $V$, iff $f_{xy}\leq F_i$. Graph edges, $E$, connect vertices that are neighbors in the density histogram, and connected components are extracted using the set union data structure with performance $O(|E| \alpha(|V|))$. Since only immediate neighbors are considered for connecting, there are $O(|V|)$ edges.

To construct the merge tree, sweeping $F_i$ from $0 \rightarrow \infty$, nodes are \textit{born} at the first $F_i$, where a new component appears. As $F_i$ is increased, the components expand until they merge with another component. When components merge, the component with the more recent birth (i.e., higher $f_{xy}$) \textit{dies}, while the component with the lower $f_{xy}$ continues. For example, in \autoref{fig:model:ct:c}, at $F_1$, the pink and purple components are about to merge. When they do at $F_2$, the pink component dies since it was born more recently (i.e., $f_{pink}>f_{purple}$), and the merged component in purple continues. Once all clusters have merged into a single component, that component dies at $\infty$ (i.e., it always exists, no matter how large $F_i$ gets).

The value of this model over the distance-based model is that it only requires the input scatterplot. It needs no information about the cluster centers, and it makes no assumptions about the distribution of points within those clusters.

\subsection{Persistence Threshold Plot}
\label{sec.model.threshold}

Thus far, the models only encode the clustering structure as a function of distance or as a function of density in the merge tree. The method to select the number of clusters that will be perceived by a user is calculated similarly, irrespective of the underlying model, though the input parameters (distance vs.\ density) have different meanings.

For this, we generate a persistence threshold plot. For a given merge tree, each component has its \textit{persistence}, $\rho$, calculated. The persistence is the difference between birth and death values of the component (i.e., $\rho=death-birth$)\footnote{For the distance-based model, birth is always $0$ making $\rho=death$. However, the full definition unifies the distance- and density-based models.}. The fundamental intuition behind persistence is that it measures the relative scale of a feature (e.g., the relative change in density), as opposed to the absolute scale of the feature (e.g., the absolute density value). \uline{We use persistence as a threshold to model the number of clusters a user would count in a scatterplot and vice versa.}

This information is represented in a \textit{persistence threshold plot} or \textit{threshold plot}. To form the plot, for the threshold $T \in [0,\infty)$, at a given $T_i$, we count the number of clusters whose $\rho>T_i$. This information is encoded into the line chart (see \autoref{fig:barcode}) by plotting the threshold $T$ horizontally and the number of clusters vertically.

Given these functions, we have the ability to determine critical thresholds (using either model) for the visual separation of clusters. For example, the red dashed lines in \autoref{fig:barcode:dens} show the persistence threshold ($T_{de}$) that corresponds to perceiving 3 clusters and vice versa. With this relationship, our models can now be used to estimate the number of clusters that a user would select in a given scatterplot.

\section{Main Experiment}
\label{sec-label:experiment}

We evaluate how well the merge tree models estimate the number of clusters perceived in a scatterplot by studying 3 factors ($S$, $N$, $P$). In addition to revisiting \ref{Hyp:H1} and \ref{Hyp:H2}, we include 3 new hypotheses:  
\begin{description}[leftmargin=!,labelindent=5pt,itemindent=-20pt]
    
    \item[{[H3]}\label{Hyp:H3}] Data point size ($P$), having a direct impact on visual density, affects the accuracy in cluster count identification in scatterplots.

    \item[{[H4]}\label{Hyp:H4}] Using a persistence threshold correlated to the distribution size~($S$) of normally distributed clusters, the \uline{distance-based model} will estimate the number of clusters perceived by users.
    
    \item[{[H5]}\label{Hyp:H5}] Using a persistence threshold correlated to the size of data point~($P$), the number of data points ($N$), and by their interaction effect ($N*P$), the \uline{density-based model} will estimate the number of clusters perceived by users.

\end{description}

\subsection{Properties and Data Generation}
\label{sec:datageneration}

Using the information learned in preliminary experiment, following values were modified for the main experiment (i.e., all others remained the same, see \autoref{sec-methodology-factors}):
\begin{itemize}[noitemsep,itemsep=4pt]

    \item \uline{Data point size/area ($P$)}: $\{1_{px}, 3_{px}, 5_{px}, 7_{px}\}$ --- 
    On the low end, $1_{px}$ point size is the minimum possible value. On the high end, $7_{px}$ was chosen in combination with the number of points to limit the maximum visual density to $\sim 30\%$ of a given stimulus. 
 
    \item \uline{Number of data points ($N$)}: $\{500,2500,12500\}$ --- 
    To decide the number of data points, we considered if data points are uniformly distributed, the \textit{maximum visual density} is $MVD=\frac{N*a}{X*Y}$, where $[X \times Y]$ are stimuli dimensions $[550 \times 550]$. With a target of $< 30\%$, using $P=7_{px}$ and $N=12500$ the visual density, $MVD=0.29$, i.e., $29\%$ of pixels filled. We noted a logarithmic effect in the preliminary experiment. Therefore, logarithmic intervals (base 5) are used.    
    
    \item \uline{Distribution size ($S$)}: $\{25_{px}, 40_{px}, 55_{px}, 70_{px}, 85_{px}\}$ --- 
    The distribution size was chosen to be similar to the preliminary experiment, slightly adjusted to have fixed intervals of $15_{px}$ between values. 
    
\end{itemize}

The data generation process is kept similar to the preliminary experiment. A key difference is that task stimuli are generated for each participant covering all combinations of factors. 
For each subject, $|S|\times|N|=15$ dataset are generated and rendered into $|15|\times|P|=60$ scatterplot stimuli. Each participant received similar variability and the same combination of factors in their stimuli.

\subsection{Study Procedure}
\label{sec:task}

This study was designed similarly to the preliminary experiment (see \autoref{sec-label:preliminary:proc}) with the following variations. Each subject was shown in stimuli from their own pool of 60 stimuli in random order, and we included a post-test questionnaire, asking participants to describe their criteria for selecting the number of clusters.

We recruited participants from Amazon's Mechanical Turk (AMT) for the IRB approved study~\cite{borgo2018information, crowston2012amazon}. Based upon a post hoc power analysis of the preliminary experiment data, we recruited a total of 40 participants (21 male, 19 female; ages: $[18-64]$, median age group: $[25-34]$) limited to the US or Canada. 45\% of participants reported having corrected vision. All participants had a HIT approval rate of $\geq 95\%$, and were compensated at US Federal minimum wage.

In total, $60$ trials $\times$ $40$ participants $= 2400$ responses were collected. We carried out some data quality checks on responses, and {the following responses were eliminated---$9$ responses with task completion time of less than 1 second and $27$ responses that ran out of time ---leaving a total of $2364$ responses for analysis.}

\para{Suitability of Studying Point Size Using AMT} 
Studying visual factors, mark size in particular, on a crowdsourced environment has potential biases due to lack of control of user hardware, retinal size, viewing distance, ambient lighting, etc. For example, search task performance decreases as the viewing angle increases~\cite{enoch1959effect}. However, this lack of control is a commonly accepted limitation in crowdsourced studies---numerous recent AMT studies have considered mark size, among other properties, that could be impacted by this lack of environmental control, e.g., Szafir's study of perceived color differences~\cite{szafir2018modeling}, Chung et al.'s evaluation of orderability in visual channels~\cite{chung2016ordered}, and Kim and Heer's study of the effectiveness of multiple visual encodings~\cite{kim2018assessing}.

\subsection{Analysis Methodology}

\begin{figure}[!t]
    \centering
    
    \begin{minipage}[t]{0.0325\linewidth}
        \rotatebox{90}{
		    \begin{minipage}{2.1cm}
		    \hfill
		    \footnotesize Threshold ($T_{di}^S$)
		    \end{minipage}
		}
	\end{minipage}
	\hspace{-7pt}
	\begin{minipage}[t]{0.54\linewidth}	
	    \centering
        \includegraphics[height=2.5cm]{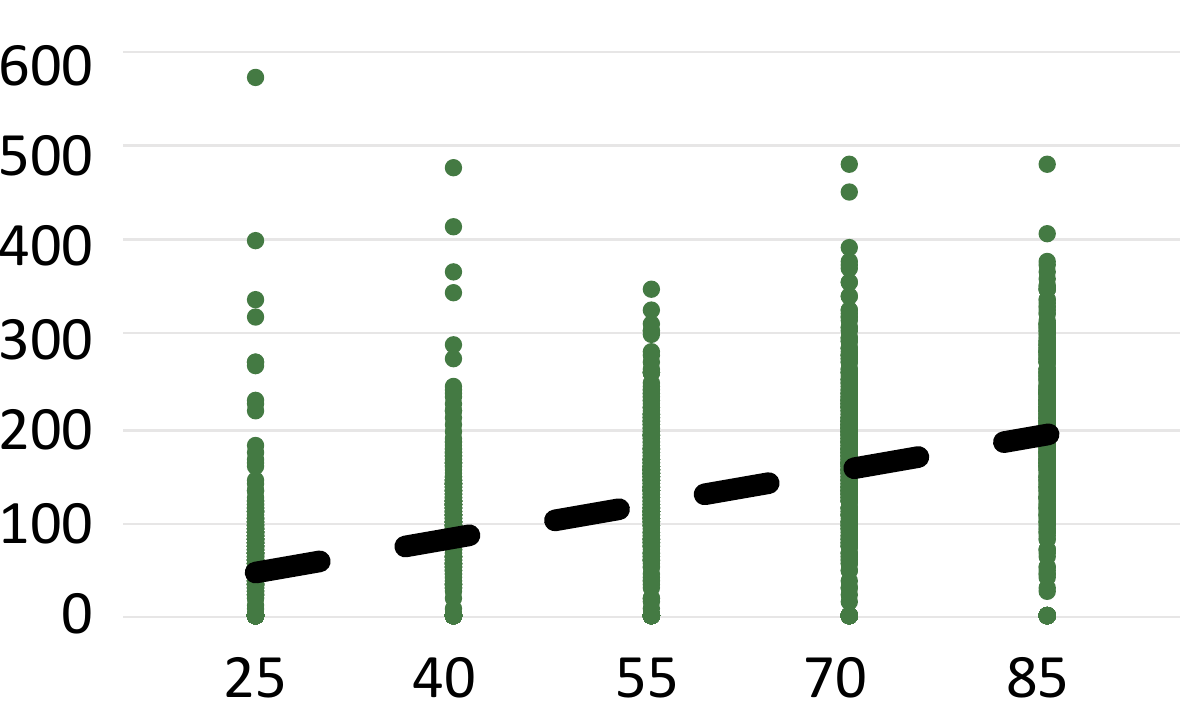}
        
	\end{minipage}
    {\begin{minipage}[t]{0.0325\linewidth}
        \rotatebox{90}{
		    \begin{minipage}{2.2cm}
		    \hfill
		    \footnotesize Differential ($\Dspace_{di}^S$)
		    \end{minipage}
		}
	\end{minipage}}	
	\hspace{-2pt}
	{\begin{minipage}[t]{0.29\linewidth}	
	    \centering
        \includegraphics[height=2.5cm]{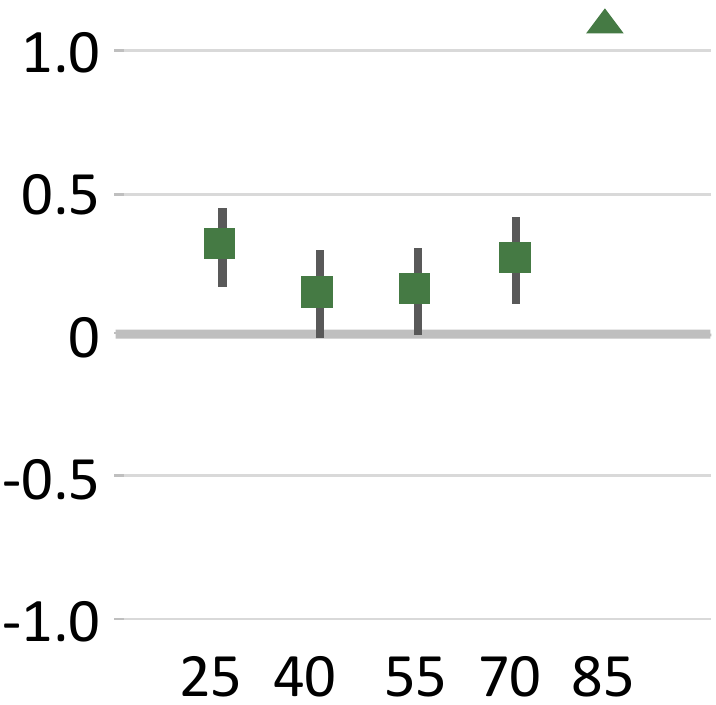}
        
	\end{minipage}}
	
	\vspace{-10pt}
	\hspace{0.16\linewidth}
    \subfigure[Threshold plot\label{fig:threshold_regression}]{\hspace{0.25\linewidth}}
    \hspace{0.2\linewidth}
    \subfigure[Differential plot\label{fig:di_dplot}]{\hspace{0.25\linewidth}}
    \hfill
	
    \caption{Plots of the threshold ($T_{di}^S$) and differential ($\Dspace_{di}^S$) against distribution size ($S$) for the distance-based model. (a) The regression (in black) $T_{di}^S(x)$ is extracted for the distribution size ($S$), in green. (b)~The regression is, in turn, used to calculate the model-predicted number of clusters, which are plotted by the mean and $95\%$ confidence interval of the differential ($\Dspace_{di}^S$) against different values of distribution size~($S$).}
    \label{fig:di_thresh_diff}
\end{figure}

We ran our data and user responses through the merge tree-based models. For the distance-based model, we first take the centers of each cluster to build the model. Then, we use the user response to the number of clusters ($U$) to extract a persistence threshold, $T_{di}$. After generating the threshold for all stimuli, a linear regression, using linear least squares, is calculated for $T_{di}$ on the factor distribution size, $T^S_{di}(s)=c_1\cdot s+c_2$, where the distribution size, $s$, is input, and $c_1$ and $c_2$ are calculated by the regression. \autoref{fig:threshold_regression} shows the resulting regression.

The density-based model is built by using the scatterplot to generate a visual density histogram, which is the input to the model. Then, the user response to the number of clusters ($U$) is used to extract a persistence threshold, $T^*_{de}$. For the density-based model, multiple factors are tested ($N$, $P$, and $N*P$), each requiring their own linear regression, i.e., $T_{de}^N(n)=c_1\cdot n+c_2$; $T_{de}^P(p)=c_1\cdot p+c_2$; and $T_{de}^{N*P}(n,p)=c_1\cdot n + c_2\cdot p+c_3$.

Threshold functions ($T^S_{di}$ and $T^*_{de}$) from the merge tree are used to calculate the \textit{model-predicted number of clusters}. To measure the accuracy of the user response on a given scatterplot, $\tau$, we add new \textit{differentials}, $\Dspace^S_{di}$ and $\Dspace^*_{de}$, for the distance- and density-based models, respectively:

\vspace{10pt}
\noindent
\begin{minipage}[t]{\linewidth}
\centering
$\Dspace_{di}^S(\tau) = U_\tau - C_{di}(T^S_{di}(\tau))$
\hspace{10pt}
$\Dspace_{de}^*(\tau) = U_\tau - C_{de}(T^*_{de}(\tau))$
\end{minipage}

\vspace{10pt}
\noindent
where $U_\tau$ is the user response, and $C_{di}$ and $C_{de}$ are the number of clusters produced by the models using a given threshold. We used the value of \textit{differentials} ($\Dspace$, $\Dspace^S_{di}$, and $\Dspace^*_{de}$) as the primary measure to analyze the effects of the factors in the cluster counting. The histograms of the differentials for both models can be found in \autoref{fig:histogram-models}.

The study followed a within-subjects design, where all 40 subjects were exposed to all the same treatment. Hence, we use repeated measures (RM) ANOVA to analyze the effects of the factors on $\Dspace$. For some results, due to violations of sphericity, according to Mauchly's test, reported degrees of freedom and $p$-values are Greenhouse-Geisser corrected (highlighted in \textcolor{green}{green})~\cite{greenhouse1959methods,mauchly1940significance}. Along with RM ANOVA, we calculated partial eta-squared ($\eta^2$). As per Cohen's guidelines for measures of $\eta^2$: $0.01$ denotes small effect, $0.06$ denotes medium effect, and $0.14$ denotes large effect~\cite{cohen1973eta}.

\subsection{Results}
\label{sec-label:result}

\newcommand{\placeImageLabelA}[1]{\put(-20,45){
  		\begin{minipage}[t][0pt][t]{0pt}
			\footnotesize
			\mbox{#1}
		\end{minipage}
		}}

\newcommand{\placeImageLabelB}[1]{\put(-26,45){
  		\begin{minipage}[t][0pt][t]{0pt}
			\footnotesize
			\mbox{#1}
		\end{minipage}
		}}

\begin{figure}[!b]
    \centering

    {\begin{minipage}[t]{0.0325\linewidth}
        \rotatebox{90}{
		    \begin{minipage}{2.35cm}
		    \centering
		    \footnotesize Frequency
		    \end{minipage}
		}
	\end{minipage}}
	\begin{minipage}[t]{0.45\linewidth}	

            \includegraphics[width=0.975\linewidth]{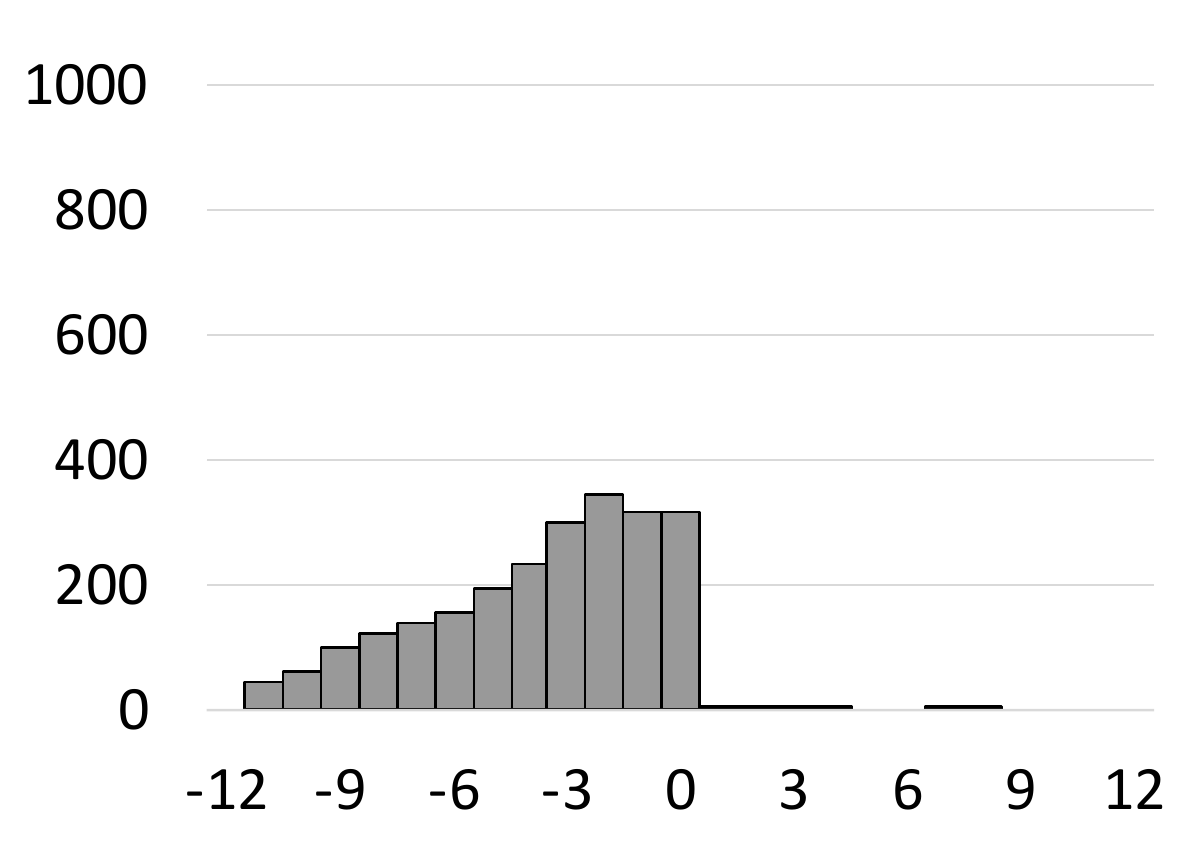}
    \end{minipage}
    \hfill
	\begin{minipage}[t]{0.45\linewidth}
            \includegraphics[width=0.975\linewidth]{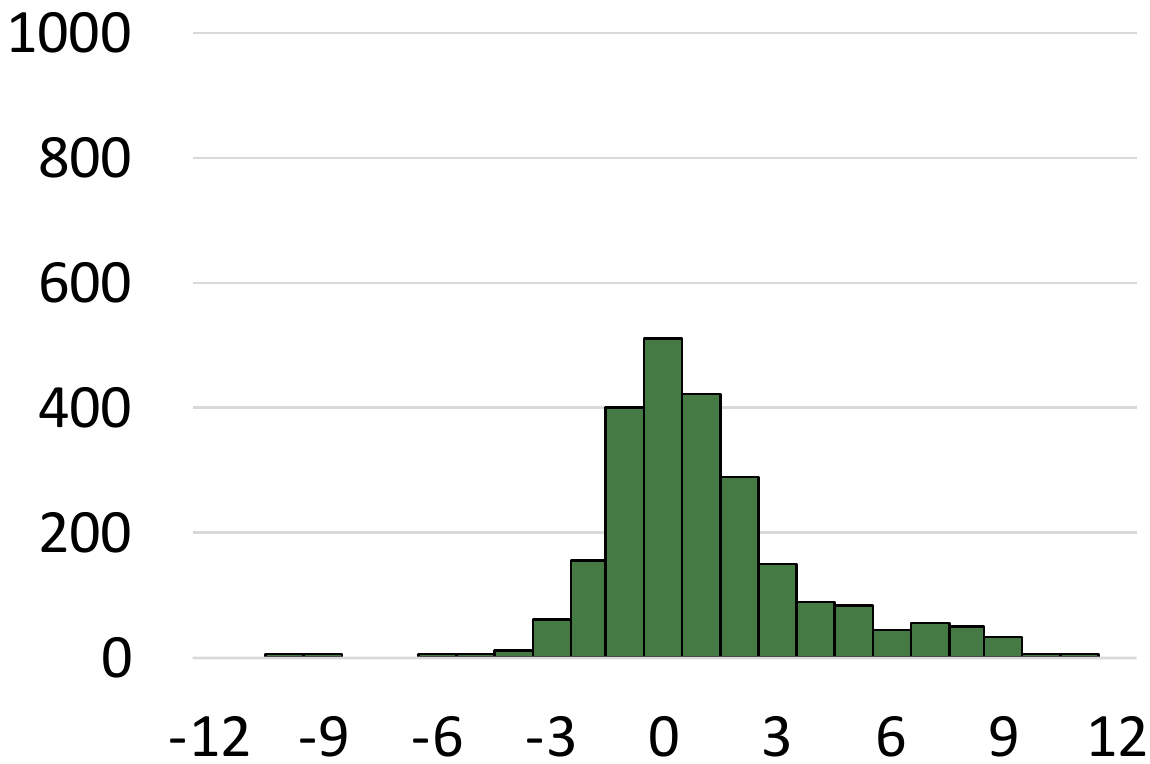}
    \end{minipage}
    \hspace{2pt}
    
    \vspace{-12pt}
    \hfill
    \subfigure[Differential ($\Dspace$) without a model\label{fig:histogram-womodel}]{\hspace{0.46\linewidth}}
    \hspace{2pt}
    \subfigure[Distance-based differential ($\Dspace_{di}^S$)\label{fig:histogram-distance}]{\hspace{0.46\linewidth}}  
    
    \subfigure[Density-based differential: number of points ($\Dspace_{de}^N$), point size ($\Dspace_{de}^P$), and interaction ($\Dspace_{de}^{N*P}$)\label{fig:histogram-density}]{
    \begin{minipage}[m]{0.98\linewidth}
        \includegraphics[width=0.32\linewidth]{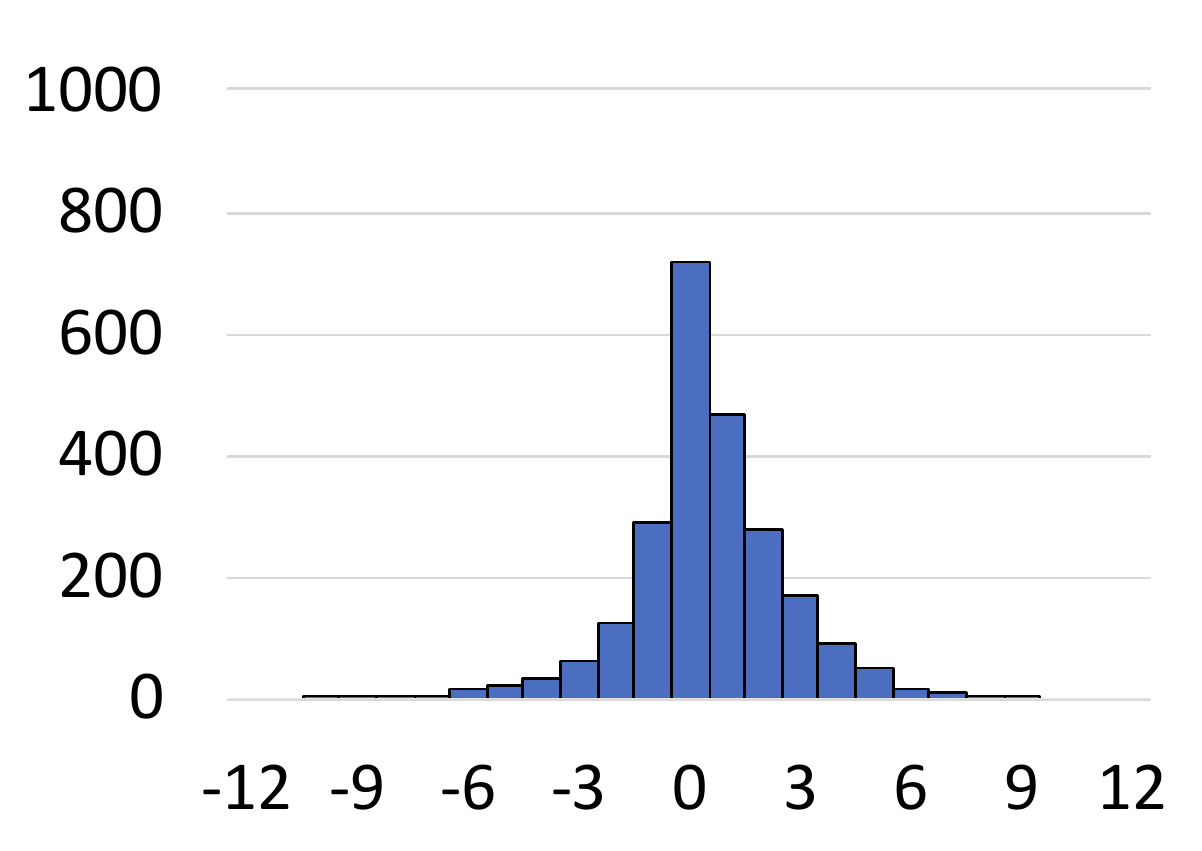}
        \placeImageLabelA{$\Dspace_{de}^N$}
        \hfill
        \includegraphics[width=0.32\linewidth]{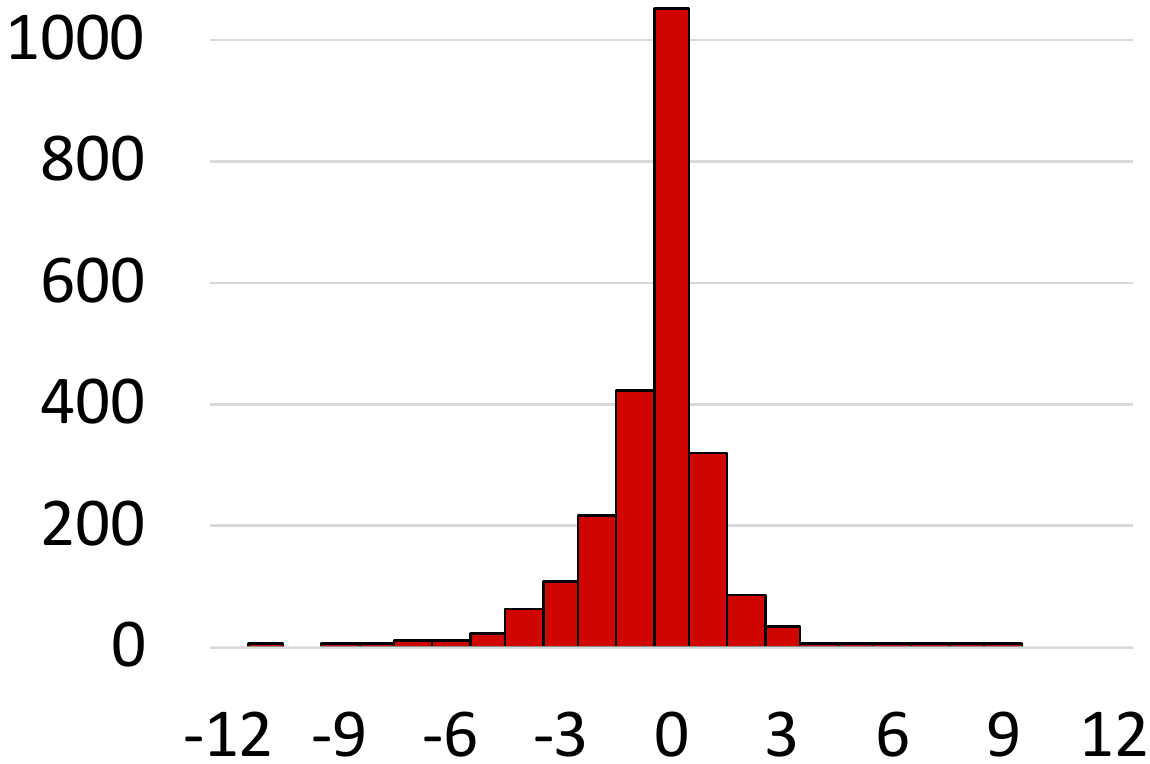} 
        \placeImageLabelA{$\Dspace_{de}^P$}
        \hfill
        \includegraphics[width=0.32\linewidth]{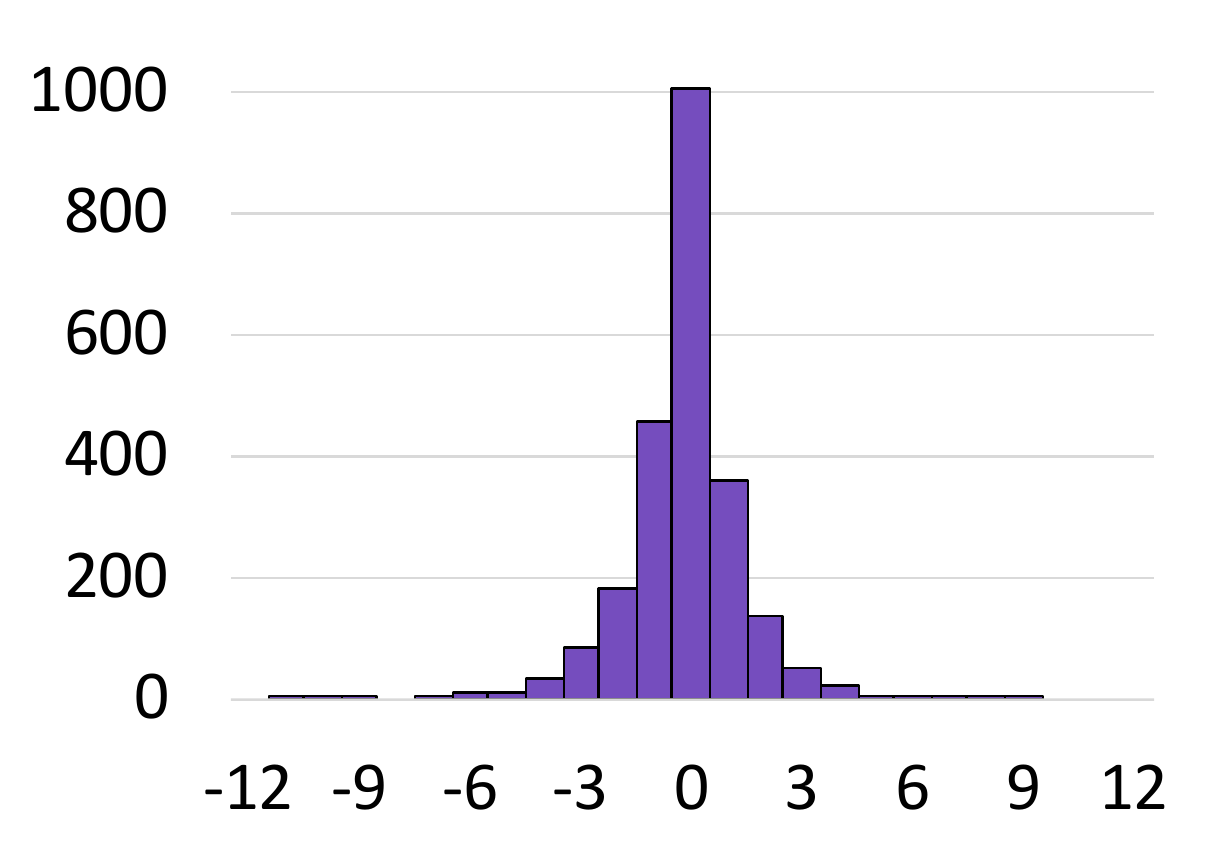}
        \placeImageLabelB{$\Dspace_{de}^{N*P}$}
    \end{minipage}
    }

    \caption{Histograms for user response {differential} (horizontally) against frequency (vertically) for (a) no model (skew due to users' underestimation), (b) the distance-based model, and (c) density-based models. Responses that are closer to $0$ imply a good fit for the data.}
    \label{fig:histogram-models}

\end{figure}

\subsubsection{Model Accuracy}
\label{sec-model-accuracy}

The distance- and density-based models both successfully estimated user perception for counting clusters. \autoref{fig:histogram-models} shows the performance of all models in terms of differential. From our analysis, \uline{we observed the highest estimation accuracy was achieved using the density-model}, from best to worst, $\Dspace_{de}^{N*P}$:~($\mu=0.18$, $\sigma=1.58$);  $\Dspace_{de}^{P}$:~($\mu=0.50$, $\sigma=1.67$); and $\Dspace_{de}^{N}$:~($\mu=-0.53$, $\sigma=2.14$). The distance-based model performs next best, $\Dspace_{di}^S$: ($\mu= 1.12$, $\sigma=2.64$). Whereas, without a model performed the worst, $\Dspace$: ($\mu=-3.74$, $\sigma=3.00$). 

\subsubsection{Factor Effect Analysis Without a Model}
\label{sec-visualfactorsanalysis}

We performed 3-factor RM ANOVA testing to analyze the factors, distribution size ($S$), number of points ($N$), and point size ($P$) in terms of the effect on the differential without a model, $\Dspace$.

We observed that the \textit{distribution size}~($S$) and the \textit{number of points}~($N$) had a significant effect for counting clusters with respect to the differential, $\Dspace$, with ($F_S(4,2304)=286.11, p< 0.001, \eta^2=0.32$) and (\textcolor{green}{$F_N(1.98,1576.43)=33.98, p<0.001$}, $\eta^2=0.029$), respectively. On the other hand, \textit{data point size} ($P$) failed to reach significance, with ($F_P(3,2304)=0.21, p = 0.889$, $\eta^2=0.0002$). We also tested for interaction effects and only observed a significant effect between $S$ and $N$, ($F_{S*N}(8,2304)=8.18, p<0.001, \eta^2=0.028$).

The $\eta^2$ analysis showed a large effect size on distribution size ($S$) and a small on the number of points ($N$) and interaction effect $S*N$. This is likely because smaller distributions create denser clusters with better separation, while larger distributions blend to create ambiguous boundaries. From these results, \textbf{both \ref{Hyp:H1} and \ref{Hyp:H2} are reconfirmed}.
The lack of significance on point size ($P$) indicates that \ref{Hyp:H3} should be rejected. However, we will revisit this hypothesis later.

\renewcommand{\arraystretch}{1.1}
\begin{table}[!t]
    \centering
    \caption{Distance-based model Repeated-Measures ANOVA results on the effect of visual factors on the \textit{differential}, $\Dspace_{di}^S$.}
    \label{tbl:factors_distance}
    \resizebox{1.0\linewidth}{!}{%
    \begin{tabular}{l || cc | cc | cc}
        Factors & Df & F-value & P-value & $\eta_{}^{2}$ & $\sigma$  \\
        \hline
        \hline

         \quad Distribution size (S) & 4 & 552.9 & $\leq 0.001$ & 0.4900 & 1.91 \\ 
         \quad Number of points (N) & \textcolor{green}{1.98} &  \textcolor{green}{051.39} &  \textcolor{green}{$\leq 0.001$} & 0.0420 & 2.61 \\    
         \quad Size of points (P) & 3 & 000.41 & 0.746 & 0.0005 & 2.64\\       
         \quad S*N & 8 & 002.72 & 0.006 & 0.0100 & - \\       
         \quad S*P & 12 & 000.13 & 1.000 & 0.0006 & -\\       
         \quad N*P & 6 & 000.08 & 0.998 & 0.0002 & -\\       
         \quad S*P*N & 24 & 000.21 & 1.000 & 0.0020 & -\\       
         \hline
    \end{tabular}} 

    \vspace{-8pt}
    \begin{flushleft}{\footnotesize 
    -: not calculated; \color{green}{Greenhouse-Geisser corrected.}}\end{flushleft}
    \vspace{-8pt}
    
\end{table}

\renewcommand{\arraystretch}{1}

\subsubsection{Distance-based Model Factor Analysis}
\label{sec-distance_analysis}

    Using persistence threshold on distribution size, $T_{di}^S$, we calculated the \textit{differential} ($\Dspace_{di}^S$) and performed 3-factor RM ANOVA  to observe the main effects of the individual factors distribution size ($S$), number of points ($N$), and point size ($P$), as well as interaction effects (see  \autoref{tbl:factors_distance}).

    The analysis identified a significant effect of distribution size~($S$) and the number of points ($N$) on the \textit{differential} ($\Dspace_{di}^S$), but the point size~($P$) failed to reach significance. 
    In particular, we found a large effect for distribution size ($S$) on $\Dspace_{di}^S$.
    We also observed a small-medium effect in the number of points ($N$) and a negligible effect on the point size ($P$). We did not anticipate any interaction effects, and only $S*N$ showed a small effect. In terms of accuracy, as pointed out in \autoref{sec-model-accuracy}, \uline{the distance-based model improved overall accuracy over using no model} (see \autoref{fig:histogram-distance}). Investigating further, \autoref{fig:di_dplot} shows the accuracy per distribution size. Note that the accuracy was sound for all distribution sizes, except at $S=85$, which negatively impacted overall performance. We speculate that this is due to the significant blending of distributions at this extreme.
    Given the large effect in $S$ and overall improvement in accuracy, \textbf{we consider \ref{Hyp:H4} confirmed}.

\begin{table*}[!ht]
        \centering
        \caption{Density-based model Repeated-Measures ANOVA results on the effect of visual factors for user response accuracy (i.e., \textit{differential}), $\Dspace_{de}^*$.}
        \label{table:tab-density-threshold}
        \resizebox{1.0\linewidth}{!}{%
        \begin{tabular}{@{\extracolsep{6pt}}c@{}c||@{}c@{}c@{}c@{}c@{ }|@{ }c@{ }c@{ }c@{ }c@{ }|@{ }c@{ }c@{ }c@{ }c@{ }|@{ }c@{ }c@{ }c@{ }c@{ }|@{ }c}
        & & \multicolumn{4}{c}{\underline{Distribution Size (S)}}& 
                     \multicolumn{4}{c}{\underline{Number of Points (N)}} & 
                     \multicolumn{4}{c}{\underline{Point Size (P)}} & 
                     \multicolumn{4}{c}{\underline{Interaction (N*P)}} \\
                 Model & Grid & Df & F-val & $p$-val & $\eta_{}^{2}$  &  Df & F-val & $p$-val & $\eta_{}^{2}$ & Df & F-val & $p$-val & $\eta_{}^{2}$ & Df & F-val & $p$-val & $\eta_{}^{2}$ & $\sigma$ \\
                 \hline \hline
                                      & $10_{px}\times10_{px}$ & 4 &  061.45 & $\leq 0.001$ & 0.090  & 2 & 122.60 & $\leq 0.001$ & 0.090  & 3 & 086.28 & $\leq 0.001$ & 0.099 & 6 &  11.28 & $\leq 0.001$   & 0.028 & 1.640\\
                \rowcolor{lightergray}
                $\Dspace_{de}^{N}$  & $20_{px}\times20_{px}$ & 4 &  014.99 & $\leq 0.001$ & 0.020  & 2 &  058.80 & $\leq 0.001$ & 0.045 & 3 & 301.80 & $\leq 0.001$ & 0.280  & 6 &  10.20  & $\leq 0.001$   & 0.025 & 1.760\\
                                      & $40_{px}\times40_{px}$ & 4 & 129.22 & $\leq 0.001$ & 0.180  & 2 & 216.00 & $\leq 0.001$ & 0.150  & 3 & 613.20 & $\leq 0.001$ & 0.430  & 6 &  56.90  & $\leq 0.001$   & 0.120  & 2.440\\
                \hline
                                      & $10_{px}\times10_{px}$ & 4 &  088.58 & $\leq 0.001$ & 0.130  & 2 &  096.50 & $\leq 0.001$ & 0.075 & 3 &   006.00 & $\leq 0.001$ & 0.086 & 6 &   1.62 & \hspace{5pt} 0.103 & 0.004 & 1.680\\
                \rowcolor{lightergray}
                 $\Dspace_{de}^{P}$   & $20_{px}\times20_{px}$ & 4 &   004.98 & $\leq 0.001$ & 0.008 & 2 & 143.80 & $\leq 0.001$ & 0.100  & 3 &  074.30 & $\leq 0.001$ & 0.080  & 6 &  11.40  & $\leq 0.001$   & 0.028 & 1.520\\
                                      & $40_{px}\times40_{px}$ & 4 &  094.45 & $\leq 0.001$ & 0.130  & 2 & 439.70 & $\leq 0.001$ & 0.270  & 3 & 135.40 & $\leq 0.001$ & 0.140  & 6 & 010.80  & $\leq 0.001$   & 0.250  & 2.070\\
                \hline
                                      & $10_{px}\times10_{px}$ & 4 &  057.95 & $\leq 0.001$ & 0.090  & 2 & 383.90 & $\leq 0.001$ & 0.240  & 3 & 246.30 & $\leq 0.001$ & 0.240  & 6 & 130.30  & $\leq 0.001$  & 0.250  & 1.710\\
                \rowcolor{lightergray}
                 $\Dspace_{de}^{N*P}$ & $20_{px}\times20_{px}$ & 4 &   007.56 & $\leq 0.001$ & 0.012 & 2 &  047.10 & $\leq 0.001$ & 0.038 & 3 &  061.20 & $\leq 0.001$ & 0.070  & 6 &  11.40  & $\leq 0.001$   & 0.028 & 1.470\\
                                      & $40_{px}\times40_{px}$ & 4 &  062.81 & $\leq 0.001$ & 0.096 & 2 & 273.30 & $\leq 0.001$ & 0.180  & 3 & 195.50 & $\leq 0.001$ & 0.190  & 6 &  19.10  & $\leq 0.001$   & 0.046 & 1.910\\
                \hline
        \end{tabular}
        }
    
    \vspace{-8pt}
    \begin{flushleft}{\footnotesize 
        \colorbox{lightergray}{The $[20_{px}\times20_{px}]$ grid is highlighted to indicate it is the primary focus of our analysis.} 
    } 
    \end{flushleft}        
    \vspace{-8pt}
    
\end{table*}

\subsubsection{Density-based Model}
\label{sec.eval.density}

    For the density-based model, we calculate 3 variations of the threshold and differential that use the factors that most directly influence visual density. Those are the {number} of data points ($T_{de}^N$/$\Dspace_{de}^N$), the {size} of data points ($T_{de}^P$/$\Dspace_{de}^P$), and their interaction ($T_{de}^{N*P}$/$\Dspace_{de}^{N*P}$). For each, we performed 3-factor RM ANOVA testing on the individual factors the distribution size ($S$), the number of points ($N$), and the point size ($P$), as well as interaction effects (see \autoref{table:tab-density-threshold}).

    \begin{figure}[!b]
        \centering
        
        \begin{minipage}[t]{0.05\linewidth}
            \rotatebox{90}{
    		    \begin{minipage}{2.5cm}
    		    \centering
    		    \footnotesize Normalized Persistence Threshold
    		    \end{minipage}
    		}
    	\end{minipage}
    	\hspace{2pt}
    	\begin{minipage}[t]{0.8\linewidth}	
    	    \centering
            \includegraphics[width=\linewidth]{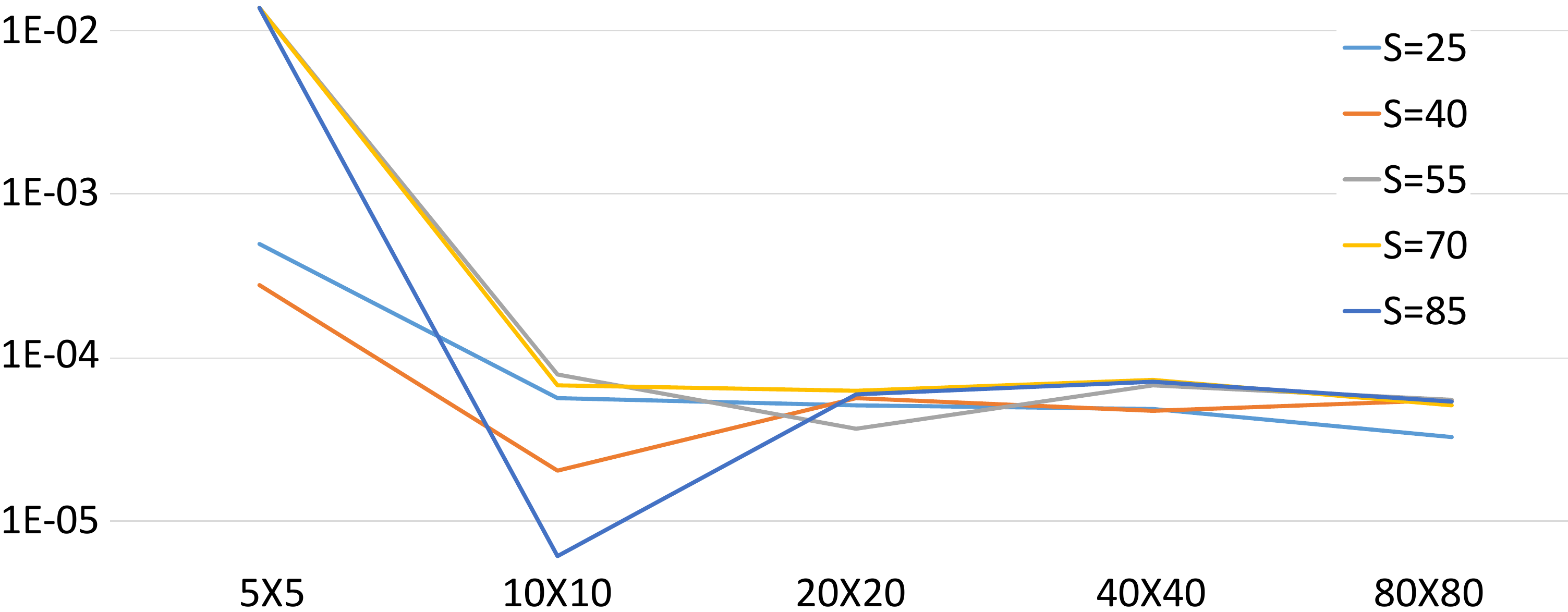}
    	\end{minipage}
    	
        \caption{Normalized persistence threshold for the density-based model for the data from \autoref{fig:example_stimuli} shows stability at histogram bins size $[20_{px}\times 20_{px}]$ and larger.}
        \label{fig:VD-plot}
    \end{figure}

    \para{Histogram Resolution} The density-based model uses the \textit{visual density} of a given scatterplot to model cluster perception. To calculate visual density, a 2D histogram is calculated on the image with bins of uniform width and height, $[B_{px}\times B_{px}]$. The choice of bin size for the density histogram is potentially influential in our analysis, as bins that are too small may cause instability, and bins that are too large may miss clusters. To determine the appropriate bin size, we performed an analysis on the data from \autoref{fig:example_stimuli}. A set of stimuli images are generated with fixed values for factors ($C=6$, $N=2500$, and $P=7_{px}$) and different values for $S=\{25_{px}, 40_{px}, 55_{px}, 70_{px}, 85_{px}\}$. We plotted the normalized density threshold (i.e., density threshold divided by area of a bin, i.e., $T_{de}/B^2$) generated by $U=6$ clusters for 5 different bin sizes (see \autoref{fig:VD-plot}). The results showed instability in the density threshold for smaller values and a stable result starting at $[20\times20]$. For this reason, 3 resolutions of histogram cell sizes are reported: $[10_{px}\times10_{px}]$, $[20_{px}\times20_{px}]$, and $[40_{px}\times40_{px}]$, but our main discussion focuses on $[20_{px}\times20_{px}]$.

\begin{figure}[!t]
    \centering
     \subfigure[$\Dspace_{de}^{N}$ for the number of points\label{fig:de_ci:a}]{\includegraphics[height=1.725cm]{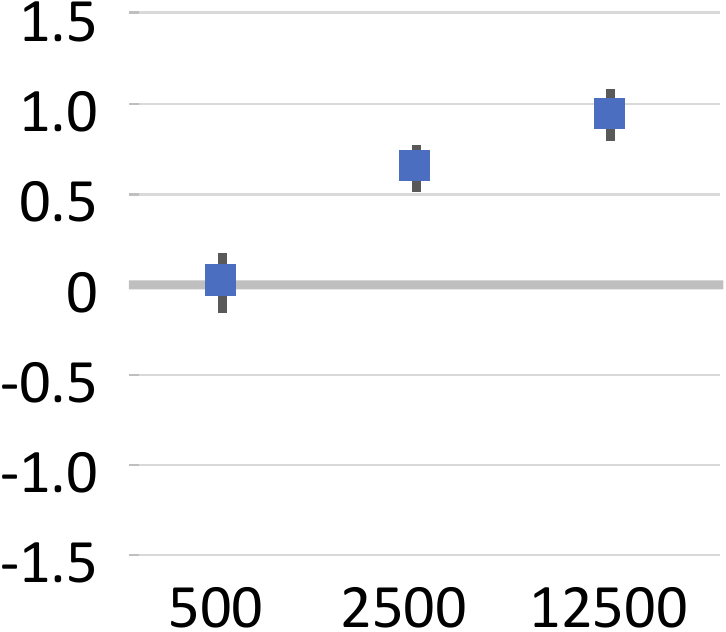}}
     \hfill
    \subfigure[$\Dspace_{de}^{P}$ for the size of points\label{fig:de_ci:b}]{\hspace{2pt}\includegraphics[trim=0pt 0 0 0, clip, height=1.725cm]{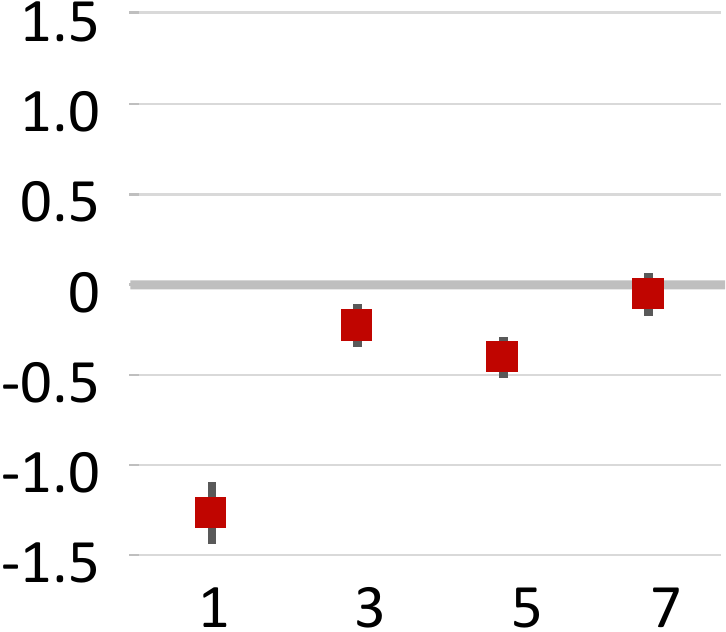}\hspace{2pt}}
    \hfill
    \subfigure[$\Dspace_{de}^{N*P}$ for the number (left) and size (right) of points.\label{fig:de_ci:c}]{\includegraphics[trim=0pt 0 0 0, clip, height=1.725cm]{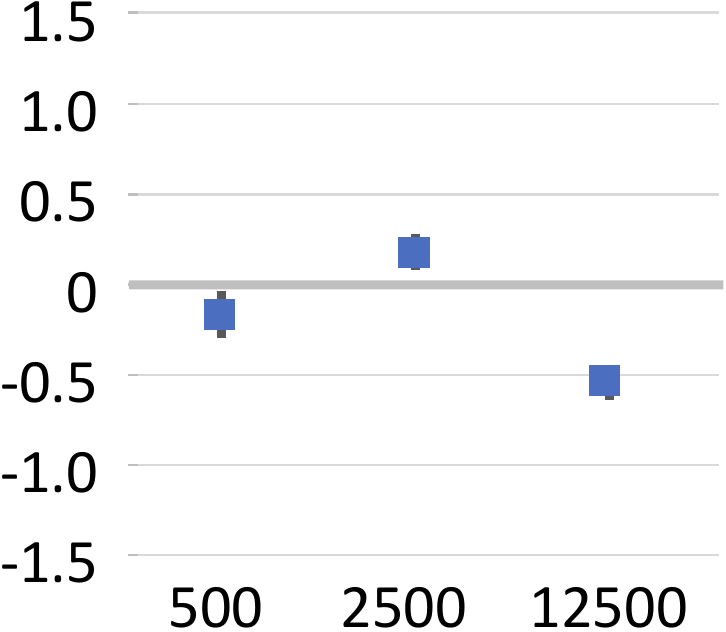}
    \hspace{3pt}
    \includegraphics[trim=0pt 0 0 0, clip, height=1.725cm]{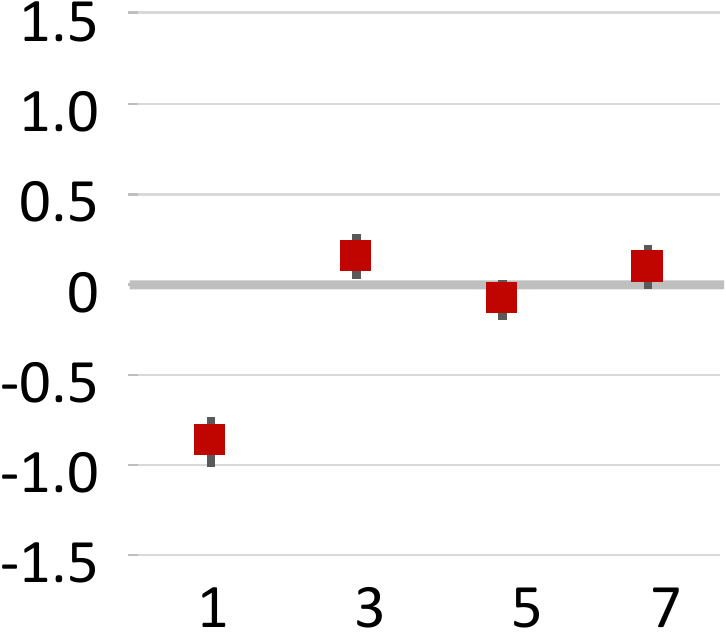}}
    \caption{Density-based model mean and $95\%$ confidence intervals of user response differential, $\Dspace_{de}^*$, on factors with density histogram $[20_{px}\times20_{px}]$.}
    \label{fig:de_ci}
\end{figure}

    \para{Number of Points Model ($T_{de}^{N}$/$\Dspace_{de}^{N}$)} 
    RM ANOVA results demonstrate significant and consistent main effects of $S$, $N$, $P$, and interaction effect of $N*P$, which can be seen in \autoref{table:tab-density-threshold}. Point size has a large effect on $\Dspace_{de}$, confirming our hypotheses and previous work (e.g., \cite{sadahiro1997cluster}) of density's influence on cluster perception. The number of points showed a small-medium effect size on $\Dspace_{de}$, also align with our hypotheses. The accuracy of the number of points model was the worst of the 3 density models, though still significantly better than no model (see \autoref{fig:histogram-density}). The accuracy of the model, plotted by the number of points in \autoref{fig:de_ci:a}, shows lower accuracy as the number of points increases.

     \para{Point Size Model ($T_{de}^{P}$/$\Dspace_{de}^{P}$)} 
     In this model, the number of points showed a medium-large effect size, while point size demonstrated a medium effect size for the differential. On the other hand, interaction of $N*P$ results small values of $\eta^2$ (see \autoref{table:tab-density-threshold}). The overall accuracy of this model was better than the number of points model (see \autoref{fig:histogram-density}). \autoref{fig:de_ci:b} shows the accuracy per point size. The model was largely accurate, except for the smallest size, $P=1_{px}$.

     \para{Interaction Model ($T_{de}^{N*P}$/$\Dspace_{de}^{N*P}$)} 
      Similar to the previous 2 models, significant effects were observed for all factors. However, only point size demonstrated medium effect size (see \autoref{table:tab-density-threshold}). \uline{This model showed the best overall accuracy of any model tested} (see \autoref{fig:histogram-density}). This makes logical sense, as the density is the combination of the number of points and their size. \autoref{fig:de_ci:c} shows the accuracy per number of points and per point size. For both cases, the accuracy was improved. However, $P=1_{px}$ was still the worst performing category.

      Our analysis showed the number of points, point size, and their interaction all had significant effects and improved accuracy over no model. \textbf{Therefore, we consider \ref{Hyp:H5} confirmed}. Furthermore, we identified some large effects with point size for the density-based model, and \textbf{this indirect relationship confirms \ref{Hyp:H3}}.

\subsubsection{Post-Test Questionnaire}
To further support our hypotheses, we asked the participants to state the criteria that influenced their counting of clusters in a free-response format at the end of the experiment. Their responses largely mirrored our findings---10\% cited the size of symbol; 25\% responses cited something amounting to distribution size; 25\% cited distance between clusters; and 65\% of responses included density as a factor\footnote{Some subjects listed multiple criteria.}.

\subsection{Follow-up Study on Opacity}
\label{sec-label:opacity}

We now evaluate opacity modification, which studies have shown to be more effective in overdraw reduction than other techniques, such as reducing point size or changing the shape of the data point~\cite{few2008solutions}. Given the results for the density-based model, we hypothesize that it will be able to model the perception of clusters when opacity is applied to the data points.
    
    \begin{description}[leftmargin=!,labelindent=5pt,itemindent=-20pt]
        \item[{[H6]}\label{Hyp:H6}] Using a density-threshold correlated to the opacity of data points ($O$), the \uline{density-based model} will have a significant effect for the number of clusters perceived by the viewer.
    
    \end{description}

\paragraph{Properties and Data Generation}
    Our data synthesis and rendering of scatterplots is similar to main experiments (see \autoref{sec:datageneration}) in most aspects. We fix the number of points $N=200,000$ and point size $P=7_{px}$ to overdraw the data on the scatterplot (see \autoref{fig:opacity_stimuli} for an example). The distribution size is the same as in the main experiment $S=\{25_{px}, 40_{px}, 55_{px}, 70_{px}, 85_{px}\}$. The data point opacity was selected on a logarithmic interval, $O=\{1\%, 10\%, 100\%\}$ over a white background. Each subject sees each condition 2 times. Thus, for each subject, $2\times|S|\times|N|=10$ datasets are generated and rendered into $|10|\times|P|\times|O|=30$ scatterplot stimuli.

\paragraph{Study Procedure}

    The task for the study was identical to the main experiment in \autoref{sec:task}. We recruited 40 participants (21 male, 19 female; median age group: $[35-44]$) from AMT, limited to subjects located in the US or Canada. 45\% of participants reported having corrected vision. Subjects were compensated at US federal minimum wage.    In total $30$ tasks $\times$ $40$ participants resulted in $1200$ responses. We carried out data quality checks on responses---2 participants ($60$ total responses) were discarded because the majority of responses were the default value of $1$, and $23$ responses that ran out of time were rejected. A total of $1117$ responses were analyzed.

\subsubsection{Analysis and Results}

    The analysis was performed similarly to \autoref{sec.eval.density}. We performed 2-factor RM ANOVA testing and evaluated the effect of factor opacity of data point ($O$) with varying the distribution size ($S$) on measure $\Dspace_{de}^O$.

   The calculation of the {visual density histogram} was modified such that it summed the pixel intensities, instead of counting the number of filled pixels. For building the histograms, we only considered the bin of size $[20_{px}\times 20_{px}]$. We calculated the {density threshold} on the opacity factor using linear least squares regression on $T_{de}^{O}(o)=c_1\cdot o+c_2$. Using $T_{de}^O$, we calculated the \textit{differential} $\Dspace_{de}^O$.

    Opacity showed a medium-large effect ($F_O(2,1102)=35.1, p< 0.001, \eta^2=0.09$), followed by a medium effect for the distribution of data points ($F_S(4,1102)=17.44, p< 0.001, \eta^2=0.086$). The interaction effect of $O$ and $S$ also showed medium-large effect ($F_{O*S}(8,1102)=12.33, p< 0.001, \eta^2=0.12$). The effects that were observed \textbf{confirm \ref{Hyp:H6}}.

    Perhaps unsurprisingly, further investigation suggested that opacity has a more substantial effect when the distribution of the data is dense and a smaller effect when the data distribution is sparse. The results confirm previous findings on scatterplot overplotting~\cite{matejka2015dynamic}, asserting that a {larger distribution size} in a scatterplot requires more opaque points, whereas a narrower distribution size requires more transparent points.

\section{Model Usage}
    
    Identifying clusters is an important low-level visual analytics task~\cite{amar2005low}, as well as in data analysis in general~\cite{otterroadmap}. Still, as mentioned in Sect.~1, clustering is an ill-posed problem, with the ``correct result'' being subject to the constraints of the algorithm or individual performing the analysis. Through our evaluation, we have shown that our models, the density-based model, in particular, performed well in estimating the number of clusters an average human would perceive. However, this in and of itself is not the real application value of the models. Instead, the models can be used to optimize the visual encodings to maximize the saliency of the visualization. Furthermore, the threshold plots provide an evidence-based rationale for design decisions.

\begin{table}[!b]
    \centering
    \caption{Summary of the main effects found in the study based on $\eta^2$. The colors are the same as in \autoref{fig:histogram-models}.}
    \label{tbl:summary}
    \includegraphics[width=0.975\linewidth]{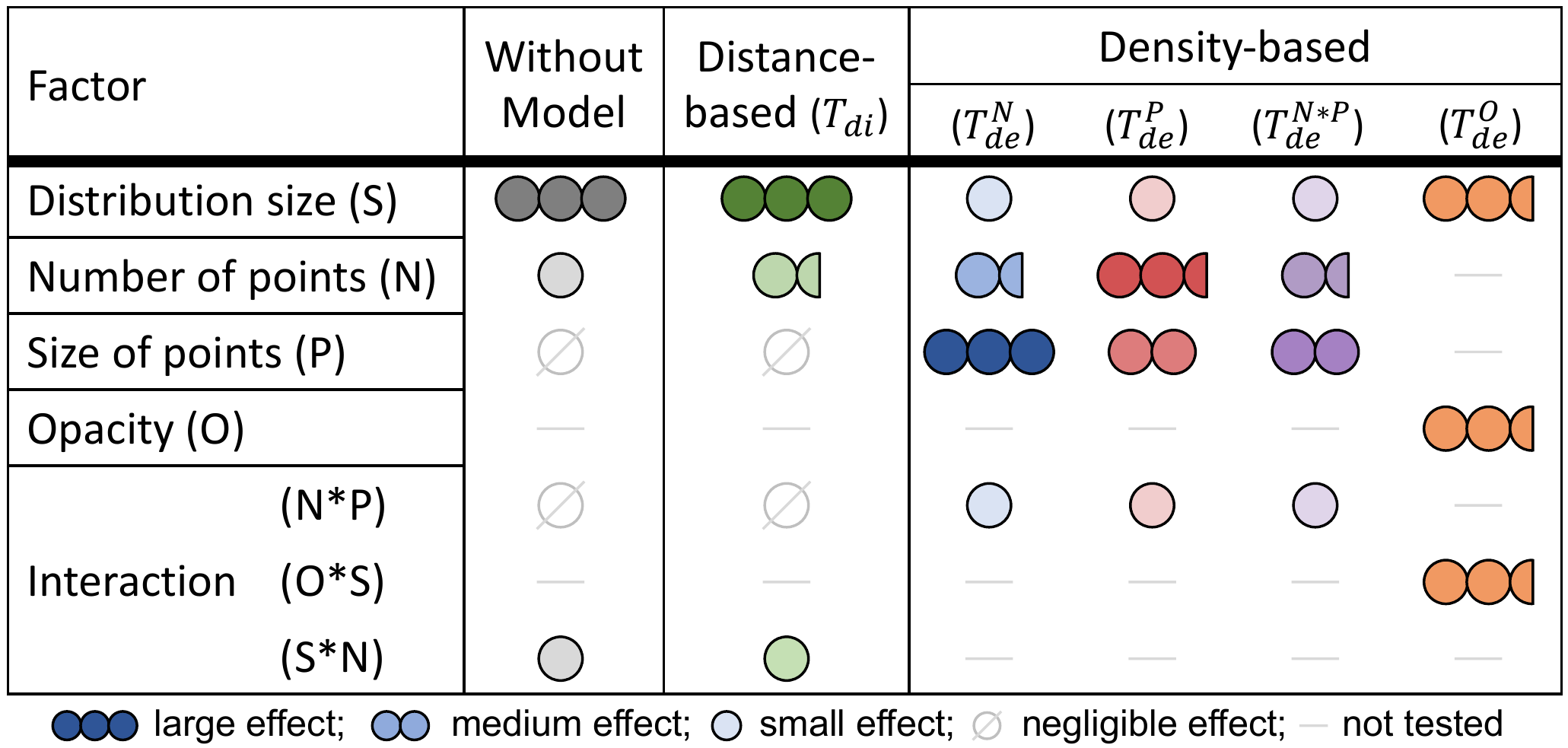}
\end{table}  

\subsection{Controlling Design Factors}
    \label{sec-modelusage:contol_factors}
    The models we have introduced construct a bridge for visualization designers between their choice of visual encodings and how users perceive clusters. \autoref{tbl:summary} summarizes our findings and suggests how designers should focus their design decisions on selecting visual properties that robustly support cluster identification in scatterplots. For the distance-based model, \textit{distribution size} ($S$) was the only factor with a large effect size. On the other hand, with the density-based model, the \textit{number of points} ($N$), \textit{point size} ($P$), and \textit{opacity} ($O$) all showed large effects on cluster count perception.

    \uline{\textit{Distribution Size}} ---
        Visualization designers generally \textit{do not} have control over {distribution size} ($S$) in the data.  Although distributions are rarely known a priori, they can be extracted from scatterplots, e.g., using Gaussian mixture models, which, combined with the distance-based model, could be used to help the designer to understand the number of clusters a user is likely to see. Nevertheless, this approach is unlikely to be helpful to the majority of designers.

    \uline{\textit{Number of Points}} ---
        Visualization designers have \textit{limited control} of the number of points ($N$), mostly in terms of data subsampling, e.g., \cite{kwon2017sampling, chen2019recursive, hu2019data}, which influences visual density.
        Using uniform random sampling, e.g.,  \cite{ellis2002density,dix2002chance}, or targeted nonuniform sampling, e.g., by using density~\cite{bertini2006give}, can reduce the number of points and, thus, the visual density. The density-based model can be used to evaluate what level of sampling provides optimal saliency of clusters.

    \uline{\textit{Point Size}} ---
        The point size ($P$) is the first design factor with \textit{complete control} in scatterplots. As pointed out earlier, increasing the area of pixels also increases the visual density. Once again, the density-based model can be used to help select the point size that provides the optimal saliency. There is an important interplay between the number of points and the point size, as adjusting either can influence the visual density.

    \uline{\textit{Opacity}} ---
        Opacity ($O$) is another factor for which the designer has \textit{complete control}, from fully transparent to fully opaque, once again impacting the visual density of the scatterplot. 
        As suggested by Urribarri and Castro, when selecting opacity, there is a trade-off with picking a point size~\cite{urribarri2017prediction} and, given our analysis, also with the number of data points shown. Nevertheless, using the density-based model, various opacity levels, along with the number of points and their size, can be evaluated and the optimal configuration selected.

\begin{figure}[!b]
    \centering

    \includegraphics[width=0.95\linewidth]{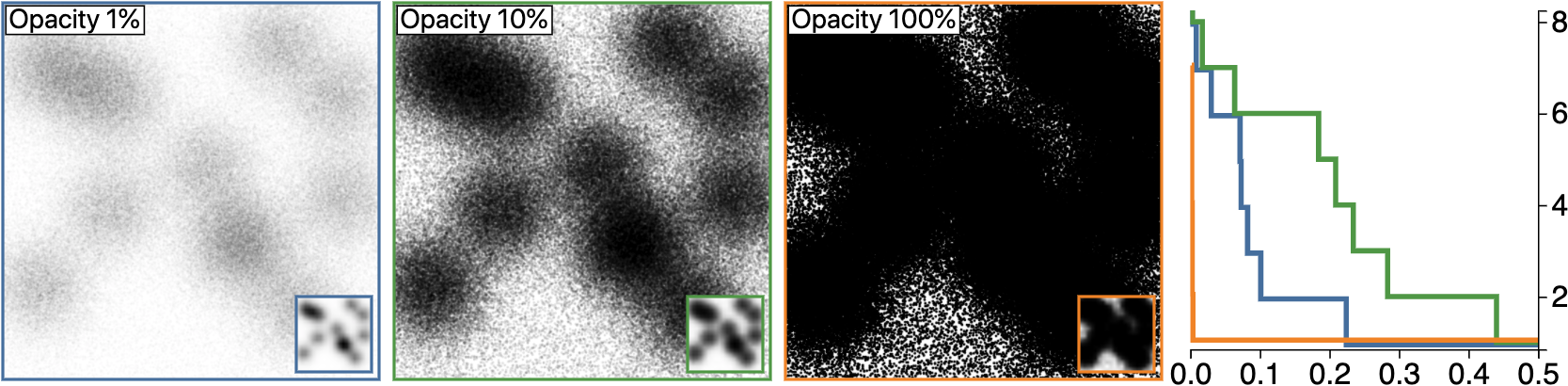}
    
	\caption{Example of overplotted stimuli with $C=11$, $N=200,000$, $S=55$, $P=7_{px}$, but varying opacity, $1\%$ (blue), $10\%$ (green), and $100\%$ (orange).} 
	\label{fig:opacity_stimuli}
 
\end{figure}

\subsection{Threshold Plots for Optimizing Cluster Saliency}

    As our goal is to improve the effectiveness of the visualization design, it is important to understand how designers can use our models to reduce ambiguity in the data, and thereby reduce the chance of misinterpretation, e.g., by having a visualization that is too sparse or over-saturated. Using pre-studies of the effects of different visual encoding configurations in scatterplots, visualization practitioners can pick the configuration that maximizes the visibility of clusters. See {\small \textless\textcolor{blue}{\url{https://usfdatavisualization.github.io/TopoClusterPerceptionDemo}}\textgreater} for a demo.
    
    Consider \autoref{fig:opacity_stimuli}, for example. The 3 scatterplots are each plotted in the persistence threshold plot. The horizontal axis reveals for each plot how salient each of the clusters are. The 10\% opacity plot shows between 1 and 8 clusters are visible, but either 2 or 6 clusters are most visible. That is not to say other numbers of clusters are not visible, but they are simply not as distinctive. Of the 3 scatterplots in \autoref{fig:opacity_stimuli}, 10\% provides the best saliency, followed by 1\%, then 100\% opacity.

\subsection{Case Study}

    To demonstrate the utility of the models on real data, we showed how the choice of visual encoding impacts the cluster perception. We performed a case study using dimensionality reduction on the MNIST dataset~\cite{lecun1998gradient}, which is an extensive database of handwritten letters commonly used to test machine learning techniques. Here we explore the dataset, which consists of 70K samples with 10 labels of handwritten digits (the labels are not used in sampling or rendering). We applied both t-SNE (see \autoref{fig:teaser:a}) and PCA (see \autoref{fig:teaser:b}) to plot the features on a 2D scatterplot and demonstrate the influence of 2 factors, the number of points and opacity.

    In \autoref{fig:teaser:a}, we show the results of varying the number of points (after dimension reduction), where $N=\{500, 2500, 12500\}$, by using random sampling. The resulting persistence threshold plot shows that for $N=500$, in red, clusters are difficult to differentiate.  For $N=2500$, in blue, and $N=12500$, in purple, both have similar levels of effectiveness, with $12500$ having a slight advantage, making it the better choice for representing this example.

    In \autoref{fig:teaser:b}, we show the results of varying the opacity of the data points, where $O=\{1\%, 5\%, 10\%, 50\%, 100\%\}$. The results in the persistence threshold plot fall into 3 groups. On the first extreme, $O=50\%$, in purple, and $O=100\%$, in orange, provide no differentiation of any clusters. On the other extreme, $O=1\%$, in blue, shows that a relatively low level of saliency for 1, 2, or 3 clusters. The final group, $O=5\%$, in pink, and $O=10\%$, in green, both show identically high levels of saliency for 1, 2, or 3 clusters in the data, making either of these the better choice for representing this example.

\section{Discussion \& Conclusions}

Scatterplots are a common type of visualization, used to identify clusters in datasets. In this work, we tested and validated the importance of 4 visual factors---{distribution size}, the {number of data points}, the {size of data points}, and the {opacity of data points}---in cluster perception and built 2 models: a distance- and density-based model for the task. Our results confirm the theoretical models of Sadahiro, which states data points distribution (proximity), and number and size of data points (concentration and density change) affect cluster perception~\cite{sadahiro1997cluster}. Finally, our findings confirm the important role that the choice of visual factors can have on cluster identification---visualization practitioners may apply these models for optimizing properties of their visualizations.

\para{Model Limitations}
Both of our models have some limitations. In the distance-based model, we required knowledge of the centers of the clusters with a fixed-size isotropic normal distribution for the model---considering other distributions would likely require modifications to the model. This requirement is particularly restrictive with respect to non-synthetic data. We showed that user response accuracy in the density-based model was significantly better than the distance-based model. However, choosing the correct density histogram resolution is a critical task that may also be dependant on the data. A choice of an extremely high or low resolution could reduce the accuracy of the threshold value. Additionally, although the density model does not directly consider a normal distribution, we have only tested it against fixed-size normal distributions. Using the model with other types of distribution should be treated with caution.

\para{Study Limitations}
The study itself has some additional limitations. First, we have not considered some other factors that could influence performance in either model, e.g., chart size, screen resolutions, etc. We have also not extensively analyzed variance between individuals, although we did note some small variation during our analysis (i.e., some individuals had over- or under-estimation tendencies).
Another limitation that we have only provided a limited analysis of mixing effects, e.g., changing the size of points, while also changing the opacity.
A final limitation is that we have not considered the correlation between confidence, which is highly related to the nature of data~\cite{etemadpour2014perception}, and the correctness on each model.

\para{Alternative Models}
Alternative models could potentially be developed to similarly explain the variance. With respect to distance, hierarchical clustering could be used, which is functionally equivalent to our distance model. For density, since stimuli are built on Gaussian distributions, a Gaussian Mixture Model (GMM) could be used. GMMs, being numerically extracted, cannot provide the same theoretical guarantees as our models, which are technically combinatorial. 
The theoretical guarantees, coming from persistent homology, also include \textit{stability} guarantees. With stability, small changes in the input are guaranteed to produce only small changes to the output. A consequence of stability is robustness to noise. The noise has low persistence, not influencing the selection of the number of clusters.

\para{Automatic Parameter Optimization}
One natural extension of this work is to develop a (semi-)automatic model for selecting design factors for a dataset. Unfortunately, using threshold plots as-is represents an under-constrained optimization, and it requires, \textit{at the very least}, a user specification of the number of clusters in the data.

\vspace{10pt}
\noindent
\resizebox{1.0\linewidth}{!}{%
\begin{minipage}[t]{1.1\linewidth}
    Code: {\small\textless\textcolor{blue}{\url{https://github.com/USFDataVisualization/TopoClusterPerception}\textgreater}}
    
Demo: {\small\textless\textcolor{blue}{\url{https://usfdatavisualization.github.io/TopoClusterPerceptionDemo}}\textgreater}
    
Data: {\small\textless\textcolor{blue}{\href{https://doi.org/10.17605/OSF.IO/49CD2}{DOI: 10.17605/OSF.IO/49CD2}}\textgreater}
\end{minipage}
}

\acknowledgments{We thank Bei Wang, Les Piegl, Jorge Adorno Nieves, Zach Beasley, Junyi Tu, and our reviewers for their input on this project. This project was supported in part by National Science Foundation (IIS-1845204).}

\bibliographystyle{abbrv-doi}

\bibliography{main}
\end{document}